\documentclass{article}[a4paper]
\usepackage[margin=1in]{geometry}
\usepackage[T1]{fontenc}

\usepackage{authblk}
\usepackage{amsfonts} 
\usepackage{amsmath}
\usepackage{amssymb}
\usepackage{amsthm}
\usepackage{mathtools}

\usepackage{hyperref}
\usepackage{xspace}
\usepackage{xcolor}

\theoremstyle{plain}
\newtheorem{theorem}{Theorem}
\newtheorem{lemma}[theorem]{Lemma}
\newtheorem{corollary}[theorem]{Corollary}
\newtheorem{fact}[theorem]{Fact}

\theoremstyle{definition}
\newtheorem{definition}[theorem]{Definition}

\title{Approximating (Weighted) Chromatic Correlation Clustering \\via Cluster LP\footnote{%
    Supported by NCN grant number 2020/39/B/ST6/01641.
    This work was partly supported by Institute of Information \& communications Technology Planning \& Evaluation (IITP) grant funded by the Korea government (MSIT) (No. RS-2021-II212068, Artificial Intelligence Innovation Hub). 
    This work was supported by the National Research Foundation of Korea (NRF) grant funded by the Korea government (MSIT) (RS-2025-00563707). 
    This work was partly supported by an IITP grant funded by the Korean Government (MSIT) (No. RS-2020-II201361, Artificial Intelligence Graduate School Program (Yonsei University)).
    Part of this work was done while H.-C.~An and C.~Lee were visiting the University of Wroc\l{}aw, Poland.
    All authors contributed equally to this work.
}%
} 

\author[1]{Fateme Abbasi}
\author[2]{Hyung-Chan An}
\author[1]{Jaros\l{}aw Byrka}
\author[2]{Changyeol Lee}
\author[1]{Yongho Shin}
\affil[1]{Institute of Computer Science, University of Wroc\l{}aw, Poland}
\affil[ ]{\texttt{\{fateme.abbasi, jby, yongho\}@cs.uni.wroc.pl}}
\affil[2]{Department of Computer Science and Engineering, Yonsei University, South Korea}
\affil[ ]{\texttt{\{hyung-chan.an, 777john\}@yonsei.ac.kr}}
\date{}

\newcommand{\cc}{\textsc{Correlation Clustering}\xspace}

\newcommand{\ccc}{\textsc{Chromatic Correlation Clustering}\xspace}
\newcommand{\wccc}{\textsc{Weighted Chromatic Correlation Clustering}\xspace}
\newcommand{\CC}{\textsc{CC}\xspace}
\newcommand{\CCC}{\textsc{CCC}\xspace}
\newcommand{\WCCC}{\textsc{WCCC}\xspace}

\newcommand{\scriptC}{\mathcal{C}}
\newcommand{\scriptCstar}{\mathcal{C}^\star}
\newcommand{\chistar}{\chi^\star}
\newcommand{\scriptCstarnear}{\mathcal{C}^\star_\mathsf{near}}
\newcommand{\chistarnear}{\chi^\star_\mathsf{near}}
\newcommand{\Cstar}{C^\star}
\newcommand{\preclustering}{\mathcal{K}}
\newcommand{\colorpreclustering}{\chi_\mathsf{pre}}
\newcommand{\initialclustering}{\mathcal{C}_\mathsf{init}}
\newcommand{\colorinitialclustering}{\chi_\mathsf{init}}

\newcommand{\epsrt}{\varepsilon_\mathrm{rt}}
\newcommand{\err}{\mathrm{err}}
\newcommand{\Eadm}{E_\mathsf{adm}}

\newcommand{\Rtilde}{\Tilde{R}}
\newcommand{\Ctilde}{\Tilde{C}}
\newcommand{\ztilde}{\Tilde{z}}
\newcommand{\xtilde}{\Tilde{x}}

\newcommand{\obj}{\mathrm{obj}}
\newcommand{\opt}{\mathrm{opt}}
\newcommand{\Nadm}{N_\mathsf{adm}}
\newcommand{\ceil}[1]{\left\lceil #1 \right\rceil}

\DeclareMathOperator{\E}{\mathbb{E}}
\DeclareMathOperator{\I}{\mathbb{I}}

\newcommand{\UBstarnear}{\textsf{UB}^\star_\textsf{near}}
\newcommand{\LBstar}{\textsf{LB}^\star}
\newcommand{\KPstarnear}{\textsf{KP}^\star_\textsf{near}}
\newcommand{\KCstarnear}{\textsf{KC}^\star_\textsf{near}}

\newcommand{\cost}{\mathsf{cost}}
\newcommand{\OPT}{\mathrm{OPT}}
\newcommand{\SOL}{\mathrm{SOL}}
\DeclareMathOperator*{\argmax}{arg\,max}

\begin{document}

\maketitle

\begin{abstract}
\cc is a fundamental clustering problem that is generalized to \ccc to incorporate categorical data. Both problems have been intensively studied, and recently, substantial improvements were obtained in the approximation algorithms for \cc. At the heart of this success lies a new linear program (LP) formulation called the \emph{cluster LP}; a natural question was whether this LP can be extended to \ccc to enable similar success.

We answer this question in the affirmative by presenting a $(2+\varepsilon)$-approximation algorithm for the problem using a \emph{chromatic} cluster LP. We then consider \wccc, in which edges have fractional weights satisfying the probability constraints, to show that our algorithm extends to this weighted version to yield the same approximation guarantee.
\end{abstract}
\section{Introduction}
Given a complete graph where each edge is labeled by either ``$+$'' or ``$-$'', \cc (\CC) asks to find a vertex clustering that minimizes the total number of \emph{disagreed edges}, where we say an edge is disagreed if it is a ``$-$'' edge within a cluster or a ``$+$'' edge across clusters.
This problem is a fundamental combinatorial optimization problem that has been extensively studied in various forms across fields~\cite{zahn1964approximating,regnie1965some,doreian1996partitioning,ben1999clustering,chen2003computing,shamir2004cluster,bansal2004correlation}.
The readers are also referred to Wahid and Hassini~\cite{wahid2022literature} for a dedicated survey on this problem.

This problem is well-known to be NP/APX-hard~\cite{shamir2004cluster,bansal2004correlation,chen2003computing,charikar2005clustering}, and recently, it is further shown that the problem cannot be approximated within a constant smaller than $24/23$ unless $\textrm{P}=\textrm{BPP}$~\cite{cao2024understanding}.
There has also been a long line of work~\cite{bansal2004correlation,charikar2005clustering,ailon2008aggregating,chawla2015near,cohen2022correlation,cohen2023handling,cao2024understanding} on improving the approximation ratio for this problem that recently culminated in $1.485$ due to Cao, Cohen-Addad, Lee, Li, Newman, and Vogl~\cite{cao2024understanding}.
A key component of their positive result is in approximately solving an exponential-sized LP, called the \emph{cluster LP}, in polynomial time and rounding this nearly optimal solution to an integral solution.

\ccc (\CCC) is a generalization of \CC motivated by clustering data with categorical interactions~\cite{bonchi2015chromatic,angel2016clustering,amburg2020clustering}.
In this generalization, ``$+$'' edges are labeled with colors. The objective is to find a clustering of vertices and a coloring of the clusters that minimize the total number of disagreements, where ``+'' edges within a cluster are deemed disagreed if their colors differ from that of the cluster.
This chromatic version has been intensively studied in the literature~\cite{bonchi2015chromatic,anava2015improved,klodt2021color,xiu2022chromatic,lee2025improved}, but previous approaches to this generalization were nevertheless limited to color-blind algorithms~\cite{klodt2021color}, reduction to the monochromatic version~\cite{anava2015improved,klodt2021color,xiu2022chromatic}, and rounding algorithms using the standard LP relaxation~\cite{anava2015improved}.
The recent work by Lee, Fan, and Lee~\cite{lee2025improved} achieving an approximation ratio of $2.15$ for \CCC is also based on reducing to monochromatic \CC using an optimal solution to the standard LP relaxation for \CCC.

In light of success of the cluster LP in \CC, it is interesting to see if a chromatic version of the cluster LP helps improve the approximation algorithm for \CCC.
In this paper, we affirmatively answer this question: we show that a natural adaptation of the cluster LP, called the \emph{chromatic cluster LP}, can be solved nearly optimally in polynomial time, and any feasible solution to the chromatic cluster LP can be rounded into an integral solution of cost within a factor of 2.
These together yield a $(2 + \varepsilon)$-approximation algorithm for \CCC, improving upon $2.15$ due to Lee et al.~\cite{lee2025improved}.
Furthermore, we identify how we can generalize this approach to the weighted case of \CCC.

In addition to \CCC, \textsc{Weighted (Monochromatic) Correlation Clustering}~\cite{bansal2004correlation,gionis2007clustering,ailon2008aggregating,van2009deterministic,chawla2015near,ostovari2025improved} is another well-studied generalization of \CC. In this problem, rather than ``integrally'' declaring an edge as ``$+$'' or ``$-$'', each edge is now assigned a fractional ``$+$'' weight and ``$-$'' weight. These two weights must be nonnegative and sum to $1$. 
This condition is thereby called the \emph{probability constraint}.
The objective of the problem is to find a clustering of vertices that minimizes the total \emph{weight} of disagreements.
It is worth noting that there exists an approximation-preserving reduction to unweighted \CC due to Chawla, Makarychev, Schramm, and Yaroslavtsev~\cite{chawla2015near}.
This reduction is based on replacing each vertex in the original weighted instance by a sufficiently large clique of ``$+$'' edges while the numbers of ``$+$'' and ``$-$'' edges between two cliques are proportional to the weights between the corresponding vertices.

In this paper, we study the weighted generalization of \CCC, called \wccc (\WCCC), which is naturally defined as follows: each edge is assigned ``$+$'' weights, one for each color, and a (colorless) ``$-$'' weight. These weights must respect the probability constraint. The goal is to find a clustering that minimizes the total weighted disagreements.
We demonstrate that our approach for \CCC smoothly extends to this weighted generalization, yielding an approximation algorithm with the same factor of $(2 + \varepsilon)$.
The aforementioned approximation-preserving reduction for the monochromatic version~\cite{chawla2015near} unfortunately does not easily extend to this chromatic version because it is unclear how one could assign colors to the edges within each clique originated from a vertex in the original instance.

Recently, Fan, Lee, and Lee~\cite{fan20261} presented a $1.64$-approximation algorithm for unweighted \CCC.
They also approximately solve the same chromatic cluster LP in polynomial time using a slightly different approach from ours, and then round this solution with both a cluster-based algorithm and a pivot-based algorithm, yielding the $1.64$-approximation algorithm.
One notable difference in solving the chromatic cluster LP lies in the construction of \emph{admissible edges} (see Section~\ref{sec:preclu:adm}): they adapted Cao et al.'s construction~\cite{cao2024understanding} into a color-sensitive version with parallel admissible edges introduced, while we show that the original construction is sufficient.
Furthermore, we extend our results to the weighted setting, demonstrating the applicability of this approach to other settings.
We note that our results for \CCC presented in this paper were obtained independently, whereas our results for \WCCC were obtained after the arXiv announcement~\cite{lee20251} of the unweighted results of Fan et al~\cite{fan20261}.
\section{Problem definitions} \label{sec_def}
We begin with defining \ccc (\CCC).
An instance of \CCC consists of a set $L$ of colors and a complete graph $(V, E)$ where each edge is specified as either  a $+$ edge or a $-$ edge, and each $+$ edge $e$ is associated with a color $c_e\in L$.
Let $E^\ell$ be the set of edges colored $\ell$; let $E^+ := \bigcup_{\ell \in L} E^\ell$ and $E^- := E \setminus E^+$ denote the sets of $+$ and $-$ edges, respectively.
Given an instance $(V, E, L,\{c_e\}_{e\in E^+})$ of the problem, a (feasible) solution consists of a partition $\scriptC$ of $V$ and a coloring function $\chi:\scriptC\to L$ that assigns a color to each part.
Given a solution $(\scriptC,\chi)$, an edge $e=uv$ is called a \emph{disagreement} if and only if one of the following conditions holds:
\begin{itemize}
    \item $e$ is a $-$ edge and both $u$ and $v$ belong to the same cluster $C \in\scriptC$;
    \item $e$ is a $+$ edge and $u$ and $v$ belong to different clusters $C\neq C'\in\scriptC$; and
    \item $e$ is a $+$ edge and both $u$ and $v$ belong to the same cluster $C \in\scriptC$, but $\chi(C)\neq c_e$.
\end{itemize}
We also refer to a disagreement simply as a \emph{disagreed} edge. If an edge is not a disagreement, we call it an \emph{agreement} or an \emph{agreed} edge.
The goal of the problem is to find a solution that minimizes the total number of disagreements, which is also referred to as the cost of the solution. 

We now turn our attention to \wccc (\WCCC).
An instance of \WCCC consists of a complete graph where each edge $e$ is associated with $|L \cup \{-\}|$ \emph{weights}, denoted by $\{w^\ell(e)\}_{\ell \in L \cup \{-\}}$.
The weights of each edge satisfy the \emph{probability constraint}, i.e., $w^\ell(e) \geq 0$ for every $\ell \in L \cup \{-\}$ and 
$
    \sum_{\ell \in L \cup \{-\}} w^\ell(e) = 1.
$
A (feasible) solution to a \WCCC instance $(V, E, L, w)$ also consists of a partition $\scriptC$ of $V$ and a coloring function $\chi:\scriptC\to L$.
To define the total cost of a solution $(\scriptC, \chi)$, we define the cost incurred by an edge $e = uv$ in the solution, denoted by $\cost_{(\scriptC, \chi)} (e)$, as follows:
\begin{itemize}
    \item $\cost_{(\scriptC,\chi)}(e):= \sum_{\ell \in L} w^\ell(e) = 1-w^-(e)$ if $u$ and $v$ belong to different clusters in $\scriptC$;
    \item
    $\cost_{(\scriptC,\chi)}(e):= \sum_{\ell \in L \cup \{-\} \setminus \{\chi(C)\} } w^\ell(e) = 1-w^{\chi(C)}(e)$ if both $u$ and $v$ belong to a cluster $C \in \scriptC$.
\end{itemize}
The cost of a solution $(\scriptC, \chi)$ is then defined to be the sum of costs incurred by all edges, i.e., $\sum_{e \in E} \cost_{(\scriptC, \chi)}(e).$
We may omit the subscript $(\scriptC, \chi)$ when it is clear from context.
The objective of the problem is to find a solution of the minimum cost.

We further define the notation that will be used.
Let $n:=|V|$ denote the number of vertices.
For the unweighted \CCC problem, let $E^{-\ell} := E \setminus E^\ell$ for every $\ell \in L$.
Given $F \subseteq E$, $u \in V$, and $S \subseteq V$, let $F(u, S) := \{uv \in F \mid v \in S\}$; for example, given $u \in V$ and $S \subseteq V$, $E^+(u,S)$ denotes the set of $+$ edges whose one endpoint is $u$ and the other endpoint is in $S$.

On the other hand, for \WCCC, given $U \subseteq L \cup \{-\}$ and $S,T\subseteq V$, let
$
    w^U(S,T):=\sum_{\ell \in U}\sum_{uv:u\in S, v\in T}w^\ell(uv).
$
For notational simplicity, let $w^U(u,T)$ and $w^U(u,v)$, respectively, denote $w^U(\{u\},T)$ and $w^U(\{u\},\{v\})$. 
For any edge $e=uv$, $w^U(u,v)$ can be written as $w^U(e)$.
Note that $w^U(u,u)=0$ by definition.
For $\ell \in L$, let $w^{-\ell}(S,T):=w^{L \cup \{-\} \setminus \{\ell\}}(S,T)$. 
Similarly, let $w^+(S,T):=w^L(S,T)$.
\section{A constant-factor approximation algorithm for \WCCC}\label{sec_constant_weighted}
This section is devoted to a constant-factor approximation algorithm for \WCCC.
The algorithm is a reduction to \CCC inspired by Bansal, Blum, and Chawla~\cite{bansal2004correlation}.
Recall that constant-factor approximation algorithms~\cite{anava2015improved,klodt2021color,xiu2022chromatic,lee2025improved} are known for \CCC.

Given an instance $I_\mathsf{ori}=(V, E, L, w_\mathsf{ori})$ of \WCCC, we construct a new instance $I_\mathsf{new}=(V, E^+ \cup E^-, L, c)$ for \CCC as follows:
\begin{align*}
    E^- 
    &
    \textstyle := \{ e \in E \mid \argmax_{\ell \in L \cup \{-\} } \{ w^\ell_\mathsf{ori}(e) \} = - \} \text{ and}
    \\
    c_e 
    & 
    \textstyle := \argmax_{\ell \in L} \{ w^\ell_\mathsf{ori}(e) \} \text{ for $e \in E^+ := E \setminus E^-$},
\end{align*}
where ties are broken arbitrarily.
We then run a $\rho$-approximation algorithm for \CCC on $I_\mathsf{new}$, and return the output of this algorithm.

Let $\cost_\mathsf{ori}(\cdot)$ and $\cost_\mathsf{new}(\cdot)$ denote the total cost of a solution with respect to $I_\mathsf{ori}$ and $I_\mathsf{new}$, respectively. 
Let $\OPT_\mathsf{ori}$ and $\OPT_\mathsf{new}$ denote optimal solutions to $I_\mathsf{ori}$ and $I_\mathsf{new}$, respectively, and let $\SOL$ denote the final output of the algorithm. 
Note that $\cost_\mathsf{new}(\SOL) \le \rho \, \cost_\mathsf{new}(\OPT_\mathsf{new})$ by the execution.

\begin{lemma}\label{const-approx-1_weighted}
    $\cost_\mathsf{ori}(\SOL)\le \cost_\mathsf{new}(\SOL) + \cost_\mathsf{ori}(\OPT_\mathsf{ori})$.
\end{lemma}
\begin{proof}
    Fix any edge $e \in E$.
    If $e$ is disagreed by $\SOL$ in $I_\mathsf{new}$, it incurs a unit cost in $I_\mathsf{new}$.
    Since the cost incurred by any edge in $I_\mathsf{ori}$ is trivially bounded by $1$, the total cost incurred by the edges disagreed by $\SOL$ is at most $\cost_\mathsf{new}(\SOL)$.
    
    On the other hand, if $e$ is agreed by $\SOL$ in $I_\mathsf{new}$, we cannot use the above argument because it incurs no cost in $I_\mathsf{new}$.
    Observe however that this edge incurs $1 - w^\ell_\mathsf{ori}(e)$ in $\SOL$ in $I_\mathsf{ori}$, where $\ell = \argmax_{\ell' \in L \cup \{-\}} \{ w^{\ell'}_\mathsf{ori}(e) \}$ by the construction of $I_\mathsf{new}$.
    Note that, in $I_\mathsf{ori}$, any solution must incur $1 - w^\ell_\mathsf{ori}(e)$ from $e$ due to the choice of $\ell$.
    Hence, the total cost incurred by the edges agreed by $\SOL$ is bounded from above by $\cost_\mathsf{ori} (\OPT_\mathsf{ori})$.
\end{proof}
\begin{lemma}\label{const-approx-2_weighted}
    For any solution $\mathcal{S}$, $\cost_\mathsf{new}(\mathcal{S}) \le 2 \, \cost_\mathsf{ori}(\mathcal{S})$.
\end{lemma}
\begin{proof}
    Fix any edge $e \in E$ disagreed by $\mathcal{S}$ in $I_\mathsf{new}$.
    Let $\ell := -$ if $e \in E^-$ and $\ell := c_e$ if $e \in E^+$.
    Observe that $w^{\ell'}_\mathsf{ori}(e) \leq \frac{1}{2}$ for any $\ell' \in L \cup \{-\} \setminus \{\ell\}$ due to the construction of $I_\mathsf{new}$. 
    Hence, the cost incurred by $e$ in $\mathcal{S}$ in $I_\mathsf{ori}$ is then at least $\min_{\ell' \in L \cup \{-\} \setminus \{\ell\}} \{ 1 - w^{\ell'}_\mathsf{ori}(e) \} \geq \frac{1}{2}$.
\end{proof}

\begin{theorem}
    This algorithm is a $(2\rho + 1)$-approximation algorithm for \WCCC.
\end{theorem}
\begin{proof}
    We can derive that
    \begin{align*}
        \cost_\mathsf{ori}(\SOL)
        & \le \cost_\mathsf{new}(\SOL) + \cost_\mathsf{ori}(\OPT_\mathsf{ori})
        \\ &
        \le \rho \, \cost_\mathsf{new}(\OPT_\mathsf{new}) + \cost_\mathsf{ori}(\OPT_\mathsf{ori})
        \\ &
        \le \rho \, \cost_\mathsf{new}(\OPT_\mathsf{ori}) + \cost_\mathsf{ori}(\OPT_\mathsf{ori}) \\ &
        \leq (2\rho+1) \, \cost_\mathsf{ori}(\OPT_\mathsf{ori}),
    \end{align*}
    where the first inequality is due to Lemma~\ref{const-approx-1_weighted}, and the last is due to Lemma~\ref{const-approx-2_weighted}.
\end{proof}

\section{Chromatic cluster LPs}\label{sec_lp}
In this section, we define the chromatic cluster LPs for \CCC and \WCCC, respectively, and present a rounding algorithm whose expected cost is at most twice the cost of the given LP solution.

\textbf{Chromatic cluster LPs.}
Define the \emph{chromatic cluster LP} for \CCC as follows:
\begin{align*}
    \text{min }
    & 
    \textstyle \sum_{S \subseteq V, \ell \in L} \left( \frac{|\delta^+(S)|}{2} + |E^{-\ell}[S]| \right) z^\ell_S \\
    \text{s.t. } 
    & 
    \textstyle \sum_{S \ni v, \ell \in L} z^\ell_S = 1, 
    && \forall v \in V, 
    \\
    &
    z^\ell_S \geq 0, 
    && 
    \forall S \subseteq V \, \forall \ell \in L,
\end{align*}
where $z^\ell_S = 1$ indicates whether $S$ is a cluster of color $\ell$, $\delta^+(S) := \{ (u,v) \in E^+ \mid u \in S, v \not\in S \}$, and $E^{-\ell}[S] := \binom{S}{2} \cap E^{-\ell}$.
Let $t^\ell_v := 1 - \sum_{S \ni v} z^\ell_S$ indicate whether $v$ is \emph{not} colored $\ell$ and $x^\ell_{uv} := 1 - \sum_{S \supseteq uv} z^\ell_S$ denote whether $u$ and $v$ are \emph{not} together in a same cluster colored $\ell$.
Observe that
\begin{align}
    \textstyle \sum_{S \ni u, S \not\ni v} z^\ell_S 
    &
    \textstyle = \sum_{S \ni u} z^\ell_S - \sum_{S \supseteq uv} z^\ell_S = x^\ell_{uv} - t^\ell_u, 
    && 
    \textstyle \forall \ell \in L \, \forall uv \in E 
    \label{eq:xminust} 
    \\
    \textstyle \sum_{\ell \in L} t^\ell_v 
    & 
    \textstyle = |L| - \sum_{\ell \in L, S \ni v} z^\ell_S = |L| - 1, 
    && 
    \forall v \in V. 
    \label{eq:sumt}
\end{align}
Using these equalities, we can rephrase the objective function as $\textstyle \obj(x) := \sum_{uv \in E^+} x^{c_{uv}}_{uv} + \sum_{uv \in E^-} \sum_{\ell \in L} (1 - x^\ell_{uv})$. We defer its derivation to Appendix~\ref{app:lp}.

The chromatic cluster LP for \WCCC has the same sets of variables and constraints, while the objective function is replaced with
\[
    \textstyle \sum_{S \subseteq V, \ell \in L} \left( \frac{w^+(S, V \setminus S)}{2} + w^{-\ell}(S, S) \right) z^\ell_S.
\]
Through a similar derivation to the unweighted version, the objective function can be rephrased as
$
    \textstyle 
    \obj(x) := 
    \sum_{uv \in E, \ell \in L} \left( w^\ell(uv) x^\ell_{uv} + w^-(uv) (1 - x^\ell_{uv}) \right).
$
The derivation can also be found in Appendix~\ref{app:lp}.

\textbf{Cluster-based algorithm.}
The algorithm is the same for \CCC and \WCCC.
Given a solution $z$ feasible to the chromatic cluster LP, consider the following algorithm.
Let $V'$ be the currently unclustered vertices (initially, $V' := V$).
Until $V'$ becomes empty, we randomly sample a cluster $S \subseteq V$ colored $\ell$ with probability $\frac{z^\ell_S}{\sum_{\ell' \in L, S' \subseteq V} z^{\ell'}_{S'}}$.
We then cluster $V' \cap S$ colored $\ell$ and update $V' := V' \setminus S$ accordingly.
This ends the description of the cluster-based algorithm.

We begin by analyzing the algorithm for \CCC.
Recall that $x^\ell_{uv} = 1 - \sum_{S \supseteq uv} z^\ell_S$ for every $\ell \in L$ and $uv \in E$.
For notational simplicity, let $x_{uv} := 1 - \sum_{\ell \in L, S \supseteq uv} z^\ell_S$ indicate whether $u$ and $v$ are \emph{not} clustered together (regardless of its color).
Note that we have $1 - x_{uv} = \sum_{\ell \in L} (1 - x^\ell_{uv})$, implying that, for any $\ell \in L$, $x_{uv} \leq x^{\ell}_{uv}$.
The following lemma is useful in showing the approximation ratios, whose proof is deferred to Appendix~\ref{app:lp}.
\begin{lemma} \label{lem:alg:key}
    For every $uv \in E$ and color $\ell \in L$, the probability that $uv$ is in a same cluster colored $\ell$ is $\frac{1 - x^\ell_{uv}}{1 + x_{uv}}$, and the probability that $uv$ is separated is $\frac{2 x_{uv}}{1 + x_{uv}}$.
\end{lemma}

\begin{theorem} \label{thm:alg}
    Given a feasible solution $z$ to the chromatic cluster LP for \CCC, the cluster-based algorithm outputs an integral solution $(\scriptC, \chi)$ whose expected cost is at most twice the objective value of $z$.
    The algorithm terminates in time $\mathsf{poly}(n, |\mathsf{supp}(z)|)$ with high probability, where $\mathsf{supp}(z)$ denotes the support set of $z$.
\end{theorem}
\begin{proof}
    From Lemma~\ref{lem:alg:key}, we can derive that the probability that $uv \in E$ is not in a cluster colored $\ell$ is at most $1 - \frac{1 - x^\ell_{uv}}{1 + x_{uv}} \leq \frac{2 x^\ell_{uv}}{1 + x_{uv}}$.
    Hence, the expected cost of $(\scriptC, \chi)$ for \CCC is bounded by
    \begin{align*}
        & 
        \textstyle \sum_{uv \in E^+} \Pr[\text{$uv$ is not in a same cluster $C$ of color $\chi(C) = c_{uv}$}]
        \\&
        \qquad \textstyle + \sum_{uv \in E^-} \Pr[\text{$uv$ is not separated}]
        \\&
        \textstyle \leq \sum_{uv \in E^+} \frac{2 x^{c_{uv}}_{uv}}{1 + x_{uv}} + \sum_{uv \in E^-} \frac{1 - x_{uv}}{1 + x_{uv}}
        \\&
        \textstyle \leq 2 \sum_{uv \in E^+} x^{c_{uv}}_{uv} + \sum_{uv \in E^-} (1 - x_{uv})
        \\&
        \textstyle \leq 2 \sum_{uv \in E^+} x^{c_{uv}}_{uv} + \sum_{uv \in E^-} \sum_{\ell \in L} (1 - x^\ell_{uv})
        \leq 2 \, \obj(x).
    \end{align*}
 
    To show that the algorithm terminates in time $\mathsf{poly}(n, |\mathsf{supp}(z)|)$, observe that, for any vertex $v \in V$, the probability of $v$ being clustered in each iteration is $\frac{\sum_{S \ni v, \ell \in L} z^\ell_S}{\sum_{\ell \in L, S \subseteq V} z^\ell_S} \geq \frac{1}{|\mathsf{supp}(z)|}$.
    Hence, using the union bound, we can show the algorithm terminates in $\Theta(|\mathsf{supp}(z)| \log n)$ iterations with high probability.
\end{proof}

For \WCCC, note that Lemma~\ref{lem:alg:key} is still satisfied because the algorithm is equivalent.
We thus have the weighted version for Theorem~\ref{thm:alg} as below.
\begin{theorem}\label{thm:alg-weighted}
    Given a feasible solution $z$ to the chromatic cluster LP for \WCCC, the cluster-based algorithm outputs an integral solution $(\scriptC, \chi)$ whose expected cost is at most twice of the objective value of $z$.
    The algorithm terminates in time $\mathsf{poly}(n, |\mathsf{supp}(z)|)$ with high probability.
\end{theorem}
\begin{proof}
    For \WCCC, observe that the objective value can be written as
    \begin{align*}
        \obj(x)
        & 
        \textstyle = \sum_{e \in E, \ell \in L} \left( w^\ell(e) x^\ell_{e} + w^-(e) (1 - x^\ell_{e}) \right)
        \\&
        \textstyle = \sum_{e \in E} \left[ \sum_{\ell \in L} w^\ell(e) x^\ell_e  + w^-(e) (1 - x_e) \right].
    \end{align*}
    It thus suffices to show that the expected cost incurred by each $e \in E$ is bounded by twice $\sum_{\ell \in L} w^\ell(e) x^\ell_{e} + w^-(e) (1 - x_{e})$.
    Observe that
    \begin{align*}
        \E[\cost(e)] 
        &
        = 
        \textstyle \sum_{\ell \in L} \Big[ w^\ell(e) \Pr[\text{$e$ separated}] + (1 - w^\ell(e)) \Pr[e \subseteq C \wedge \chi(C) = \ell] \Big]
        \\&
        \leq
        \textstyle \sum_{\ell \in L} \Big[ w^\ell(e) \, 2x_e + (1 - w^\ell(e)) \, (1 - x^\ell_e) \Big]
        \\&
        \textstyle = \sum_{\ell \in L} \Big[ w^\ell(e) (x_e + x^\ell_e) + (1 - x^\ell_e) - w^\ell(e) (1-x_e) \Big]
        \\&
        \textstyle = \sum_{\ell \in L} w^\ell(e) (x_e + x^\ell_e) + w^-(e) (1 - x_e)
        \\&
        \textstyle \leq 2 \sum_{\ell \in L} w^\ell(e) x^\ell_e + w^-(e) (1 - x_e),
    \end{align*}
    where the first inequality follows from Lemma~\ref{lem:alg:key}, the last equality from the fact that $w^-(e) = 1 - \sum_{\ell \in L}w^\ell(e)$ and $1 - x_e = \sum_{\ell \in L} (1-x^\ell_e)$, and the last inequality from the fact that $x_e \leq x^\ell_e$ for any $\ell \in L$.

    The same argument used in the proof of Theorem~\ref{thm:alg} for termination completes the proof.
\end{proof}

\section{Preclustering} \label{sec:preclu}
Due to Theorems~\ref{thm:alg} and \ref{thm:alg-weighted}, to obtain a $(2 + \varepsilon)$-approximation algorithm, it suffices to argue that a nearly optimal solution to the chromatic cluster LPs can be found in polynomial time.
To this end, we adapt the notion of a \emph{preclustered instance}~\cite{cohen2023handling,cao2024understanding} to the chromatic versions as follows.
\begin{definition}[Preclustered instance]
    Given an instance, a \emph{preclustered instance} is defined by a triple $(\preclustering, \colorpreclustering, \Eadm)$ such that
    \begin{itemize}
        \item $\preclustering$ is a partition of $V$;
        \item $\colorpreclustering : \preclustering_{>1} \to L$ is a color assignment for every non-singleton part in $\preclustering$, where we denote by $\preclustering_{>1} := \{K \in \preclustering \mid |K| > 1\}$ the set of non-singleton parts of $\preclustering$; and
        \item $\Eadm \subseteq \{uv \in \binom{V}{2} \mid \not\exists K \in \preclustering \text{ such that } uv \subseteq K \}$ is a set of edges across $\preclustering$.
    \end{itemize}
    We call $\preclustering$ and $\Eadm$ a \emph{preclustering} and \emph{admissible} edges, respectively.
    Each part in $\preclustering$ is called a \emph{precluster}.
    Any pair $uv$ of two vertices $u$ and $v$ in a same precluster is called an \emph{atomic} edge.
    Any pair that is neither atomic nor admissible is called a \emph{non-admissible} edge.
\end{definition}

\begin{definition} [Respecting solution] \label{defn:respect}
    Given a preclustered instance $(\preclustering, \colorpreclustering, \Eadm)$ and a feasible solution $(\scriptC, \chi)$, we say the solution \emph{respects} the preclustered instance when
    \begin{itemize}
        \item $\preclustering$ subdivides $\scriptC$, i.e., for every precluster $K \in \preclustering$, there exists a cluster $C \in \scriptC$ such that $K \subseteq C$;
        \item for every non-singleton precluster $K \in \preclustering_{>1}$ and the cluster $C \in \scriptC$ containing $K$, $\colorpreclustering(K) = \chi(C)$; and
        \item for every non-admissible edge $uv$, the two vertices $u$ and $v$ are not in a same cluster in the solution.
    \end{itemize}
\end{definition}

In this section, we prove the following theorem.
\begin{theorem} \label{thm:preclu:main}
    For any sufficiently small constant $\varepsilon > 0$, there exists an algorithm that, in time $n^{\mathsf{poly}(\varepsilon^{-1})}$, outputs a preclustered instance $(\preclustering, \colorpreclustering, \Eadm)$ such that there exists a solution respecting the preclustered instance whose cost is at most $(1 + \varepsilon) \, \opt$, and $|\Eadm| \leq O(\varepsilon^{-2}) \, \opt$, where $\opt$ denotes the cost of an optimal solution in the instance input to the algorithm.
\end{theorem}

\subsection{Constructing \texorpdfstring{$\preclustering$}{K} and \texorpdfstring{$\colorpreclustering$}{chi pre}} \label{sec:preclu:K}
In this subsection, we present an algorithm for constructing a preclustering $\preclustering$ and a color assignment $\colorpreclustering : \preclustering_{>1} \to L$.
The algorithm is an adaptation of Cao et al.'s construction~\cite{cao2024understanding} to the chromatic versions.
In particular, one subtlety in our algorithm is that we do not assume a self-loop for each vertex which is no loss of generality in the monochromatic version.

\textbf{Algorithm description.}
We first describe the algorithm for \CCC, and then state the changes needed for \WCCC.
We use parameters $\alpha,\beta\in(0,1)$ such that $\eta:=\frac{\alpha+\beta}{1-\beta}\in (0,\frac{1}{13})$. 
Note that $\alpha=\beta=0.02$ is one feasible choice.\footnote{
The choice of $\alpha$ and $\beta$ affects the constant hidden in the exponent of the running time $n^{\mathsf{poly}(\varepsilon^{-1})}$ for constructing a nearly optimal solution to the chromatic cluster LPs: this constant decreases as $\alpha\beta$ increases.
}

At the very start, we obtain a solution $(\initialclustering,\colorinitialclustering)$ by running any constant-factor approximation algorithm for \CCC on the given instance. 
We then iterate over every vertex $u \in V$ in an arbitrary order, and mark $u$ if either
\begin{equation} \label{eq:preclu:markcond}
    |E^{-\ell}(u,C)| \ge \alpha (|C|-1)
    \text{ or }
    |E^+(u,V\setminus C)|\ge \alpha (|C|-1) 
    \text{ (or both),}
\end{equation}
where we denote by $C \in \initialclustering$ the cluster containing $u$ and by $\ell := \colorinitialclustering(C)$ its color.
After marking all such vertices, we now iterate over every cluster $C\in \initialclustering$, and 
if the number of marked vertices in $C$ is at least $\beta (|C| - 1)$, we mark all the vertices in $C$.
We obtain $\preclustering$ by making all marked vertices singleton clusters from $\initialclustering$. The color assignment $\colorpreclustering$ is constructed by inheriting $\colorinitialclustering$, i.e., for any non-singleton $K \in \preclustering_{>1}$, we have $\colorpreclustering(K)=\colorinitialclustering(C)$, where $C \supseteq K$ is the cluster in $\initialclustering$ containing $K$. 

For \WCCC, we only need to change the condition \eqref{eq:preclu:markcond} of $u \in V$ being marked in the first iteration to
\begin{equation*} 
    w^{-\ell}(u,C) \ge \alpha (|C|-1)
    \text{ or }
    w^+(u,V\setminus C) \ge \alpha (|C|-1)
    \text{ (or both).}
\end{equation*}
Recall that we present a constant-factor approximation algorithm for \WCCC in Section~\ref{sec_constant_weighted}.

\textbf{Sketch of analysis.}
Let us now sketch the analysis of this algorithm.
We focus on \CCC while we mention the changes needed for \WCCC at the end; their full analyses are deferred to Appendices~\ref{app:preclu:K} and \ref{app:preclu:K_weighted}, respectively.

We aim at proving the first two conditions of Definition~\ref{defn:respect} are satisfied by any optimal solution with the constructed $\preclustering$ and $\colorpreclustering$.
Fix an arbitrary optimal solution $(\scriptCstar, \chistar)$.
The next two lemmas are useful in the proof of the main lemmas of this subsection.
\begin{lemma}\label{lem-main:precluster-disagreement-bound}
    For any non-singleton precluster $K \in \preclustering_{>1}$ with $\ell:=\colorpreclustering(K)$ and $u \in K$, we have
    $|E^{-\ell}(u,K)| < \eta (|K|-1)$ and $|E^+(u,V\setminus K)|< \eta (|K|-1)$.
\end{lemma}
\begin{lemma}\label{lem-main:precluster-domination}
    For any non-singleton precluster $K\in\preclustering_{>1}$, if a cluster $\Cstar_1\in\scriptCstar$ colored $\colorpreclustering(K)$ intersects $K$ (i.e., $\chistar(\Cstar_1) = \colorpreclustering(K)$ and $K \cap \Cstar_1 \neq \emptyset$), we have $|\Cstar_1 \setminus K| < 2\eta |K|$.
\end{lemma}
\noindent Intuitively speaking, the former lemma follows from the execution of the algorithm in a fairly straightforward manner.
The latter can be derived because, otherwise, splitting $\Cstar_1$ into $\Cstar_1 \setminus K$ and $\Cstar_1 \cap K$ in $\scriptCstar$ incurs less cost due to the second condition of Lemma~\ref{lem-main:precluster-disagreement-bound}.

The following is the main technical lemma for the first condition of Definition~\ref{defn:respect}.
We provide a proof sketch of this lemma while the full proof can be found in Appendix~\ref{app:preclu:K}.
\begin{lemma}\label{lem-main:precluster-subdivide}
    $\preclustering$ subdivides $\scriptCstar$, i.e., for any $K\in\preclustering$, there is $\Cstar \in \scriptCstar$ such that $K \subseteq \Cstar$.
\end{lemma}
\begin{proof}[Proof sketch]
    Suppose toward contradiction that $\scriptCstar$ is not subdivided by $\preclustering$.
    Then, there exists a non-singleton precluster $K \in \preclustering_{>1}$ such that $K \setminus \Cstar$ is nonempty for every cluster $\Cstar \in \scriptCstar$.
    We choose an arbitrary such precluster $K \in \preclustering_{>1}$, and let $\ell:=\colorpreclustering(K)$ be the color of $K$ in the preclustering.
    Let $\xi$ be a constant between $\frac{1+5\eta}{2}$ and $1-4\eta$.
    Note that we can always choose such $\xi$ because $\frac{1+5\eta}{2}< 1-4\eta$ for $\eta \in (0,\frac{1}{13})$.
    Observe that $\frac{1}{2}<\xi< 1$.

    We break the proof into two cases depending on the existence of a cluster $\Cstar \in \scriptCstar$ that ``dominantly'' intersects $K$, i.e., $|K \cap \Cstar| \geq \xi |K|$.
    In each case, we respectively construct a new solution whose cost is strictly less than $(\scriptCstar, \chistar)$, leading to contradiction that $(\scriptCstar, \chistar)$ is optimal.

    \textbf{Case 1. No dominant intersections.} Suppose that, for all $\Cstar \in \scriptCstar$, $|K \cap \Cstar| < \xi |K|$.
    In this case, we consider a new solution $(\scriptC, \chi)$ by removing $K \cap \Cstar$ from each $\Cstar \in \scriptCstar$ and inserting $K$ colored $\ell := \colorpreclustering(K)$.
    
    The ``gain'' of $(\scriptC, \chi)$ over $(\scriptCstar, \chistar)$ is at least the total number of $\ell$-colored edges within $K$ but crossing $\scriptCstar$.
    For each $u \in K$, at least $(1 - \xi) |K|$ edges are crossing $\scriptCstar$ inside $K$ due to the condition of this case, while at most $\eta |K|$ edges can be colored other than $\ell$ due to Lemma~\ref{lem-main:precluster-disagreement-bound}, implying that the gain is at least $\frac{1 - \xi - \eta}{2} |K|^2$.
    
    On the other hand, the ``loss'' of $(\scriptC, \chi)$ over $(\scriptCstar, \chistar)$ is due to either the appropriately colored $+$~edges crossing $K$ but within a cluster in $\scriptCstar$ or the $-$~edges within $K$ but crossing $\scriptCstar$. 
    By Lemma~\ref{lem-main:precluster-disagreement-bound}, we can bound this loss from above by $\frac{3\eta}{2} |K|^2 < \frac{1 - \xi - \eta}{2} |K|^2$, where the inequality is due to the choice of $\xi$.

    \textbf{Case 2. A dominant intersection exists.}
    Suppose now there exists a cluster $\Cstar_1 \in \scriptCstar$ such that $|K \cap \Cstar_1| \ge \xi |K|$.
    Note that we always have at most one such cluster since $\xi>\frac{1}{2}$.
    Let $K_1:=K \cap \Cstar_1$.
    
    Assume for now that $\chistar(\Cstar_1) = \colorpreclustering(K)$. 
    We construct a solution $(\scriptC,\chi)$ by transferring one arbitrary vertex $u \in K \setminus \Cstar_1$ to $\Cstar_1$.
    We can then see that the gain of $(\scriptC, \chi)$ over $(\scriptCstar, \chistar)$ is due to the $\ell$-colored edges between $u$ and $\Cstar_1$, whereas the loss is due to the appropriately colored $+$~edges incident with $u$ within the cluster containing $u$ in $\scriptCstar$ and edges other than $\ell$-colored ones between $u$ and $\Cstar_1$.
    Using the dominance condition of this case and Lemma~\ref{lem-main:precluster-disagreement-bound}, we can show that the gain is at least $(\xi - \eta) |K|$.

    To bound the loss from above, we prove that the number of the first type of edges is at most $(1-\xi + \eta) |K|$ using the fact that $|K \setminus \Cstar_1| \leq (1 - \xi)|K|$ due to the case condition, and Lemma~\ref{lem-main:precluster-disagreement-bound}.
    On the other hand, the second type of edges can be partitioned into ones within $K$ and ones incident with $\Cstar_1 \setminus K$; the former is bounded by $\eta |K|$ due to Lemma~\ref{lem-main:precluster-disagreement-bound} while the latter is bounded by $2\eta |K|$ due to Lemma~\ref{lem-main:precluster-domination}.
    Hence, the loss cannot exceed $(1 - \xi + 4\eta)|K| < (\xi - \eta)|K|$ by the choice of $\xi$.

    Finally, for the case where $\chistar(\Cstar_1) \neq \colorpreclustering(K)$, we construct a solution $(\scriptC, \chi)$ from $(\scriptCstar, \chistar)$ by splitting $\Cstar_1$ into $K_1 := K \cap \Cstar_1$ colored $\ell = \colorpreclustering(K)$ and $\Cstar_1 \setminus K_1$ colored $\ell' :=\chistar(\Cstar_1)$. 
    We observe that the gain of $(\scriptC, \chi)$ against $(\scriptCstar, \chistar)$ is at least the number of the $\ell$-colored edges within $K_1$ while the loss is at most the number of the $\ell'$-colored edges within $K_1$ and between $K_1$ and $\Cstar_1 \setminus K_1$.
    Through a similar argument using Lemma~\ref{lem-main:precluster-disagreement-bound}, we can show that the gain is at least $\frac{1}{2}|K_1|(\xi|K|-1 -\eta(|K|-1))$ while the loss is at most $\frac{3}{2}\eta|K_1|(|K| - 1)$.
    The net gain is therefore at least $\frac{1}{2}|K_1|\cdot(\xi|K|-1 -4\eta(|K|-1)) > 0$, where the inequality can be derived from the fact that $K$ is non-singleton and the choice of $\xi$ and $\eta$.
\end{proof}

The second condition of Definition~\ref{defn:respect} is also satisfied through a similar argument.
The proof can be found in Appendix~\ref{app:preclu:K}.
\begin{lemma}\label{lem-main:precluster-unique}
    For every cluster $\Cstar \in \scriptCstar$ and any non-singleton precluster $K \in \preclustering$ contained in $\Cstar$, we have $\colorpreclustering(K) =  \chistar(\Cstar)$.
\end{lemma}

For \WCCC, we generalize the above analysis for \CCC by substituting the number of edges satisfying a condition with the total sum of edge weights satisfying the same condition.
In particular, instead of Lemma~\ref{lem-main:precluster-disagreement-bound} for \CCC, we have the following lemma for \WCCC: 
\begin{lemma}\label{lem-main:precluster-disagreement-bound-weighted}
    For any non-singleton precluster $K \in \preclustering_{>1}$ with $\ell:=\colorpreclustering(K)$ and $u \in K$, we have
    $w^{-\ell}(u,K) < \eta (|K| - 1)$ and $w^+(u,V \setminus K) < \eta (|K| - 1)$.
\end{lemma}
\noindent Likewise, we have the counterparts of Lemmas~\ref{lem-main:precluster-domination}, \ref{lem-main:precluster-subdivide}, and \ref{lem-main:precluster-unique} for \WCCC. The full analysis is deferred to Appendix~\ref{app:preclu:K_weighted}.

\subsection{Constructing \texorpdfstring{$\Eadm$}{E adm}} \label{sec:preclu:adm}
We now present the construction of admissible edges $\Eadm$.
Interestingly, it turns out that a color-oblivious construction as in Cao et al.~\cite{cao2024understanding} suffices for the chromatic versions.
Here we present its construction in the notation of \WCCC.

For the unweighted version, it is noteworthy that the original construction~\cite{cao2024understanding} for the unweighted monochromatic version already incorporates weights.
Hence, there exists a straightforward interpretation for \CCC where we define $w^+(S,S'):=|\{uv \in E^+ : u\in S,v\in S' \}|$ and $w^-(S,S'):=|\{uv \in E^- : u\in S,v\in S' \}|$ for $S, S' \subseteq V$.

Given $\preclustering$ from the previous section, the construction of $\Eadm$ is as follows.
For $K \in \preclustering$, let $d(K):=\frac{w^+(K,V\setminus K)}{|K|}+\frac{|K|}{2}$, $N_1(K):=\{K' \in \preclustering \mid \varepsilon \cdot d(K') < d(K) < \frac{1}{\varepsilon} \cdot d(K')\}$, and $ N_2(K):=\{K' \in \preclustering \setminus \{K\} \mid K,K' \in N_1(K)\cap N_1(K')\}$. 
For distinct $K,K' \in \preclustering$ and $\preclustering'\subseteq \preclustering$ such that $K,K' \in \preclustering'$, we further define
\begin{align*}
    W(K,K',\preclustering')
    & 
    \textstyle :=\sum_{K'' \in \preclustering' \setminus \{K,K'\}} \left[ |K''|\cdot\frac{w^+(K,K'')}{|K|\cdot |K''|}\cdot \frac{w^+(K',K'')}{|K'|\cdot |K''|} \right]
    \\&
    \textstyle \qquad +\frac{w^+(K,K')}{|K|}+\frac{w^+(K',K)}{|K'|}
\end{align*}
Moreover, for two distinct preclusters $K, K' \in \preclustering$, we denote by $K \sim K'$ if $K'\in N_2(K)$ and $W(K,K',N_1(K)\cap N_1(K'))>\varepsilon(d(K)+d(K'))$.
We can then define $\Eadm :=\{ uv\in E \mid  \exists K,K'\in\preclustering \text{ such that } K\sim K', u\in K, v\in K' \}$.

We show in Appendix~\ref{app:preclu:adm} that this $\Eadm$, together with $\preclustering$ and $\colorpreclustering$ from the previous section, satisfies the conditions of Theorem~\ref{thm:preclu:main}.

\section{Solving the chromatic cluster LP}\label{sec_sampling}
In this section, we present an algorithm for solving the chromatic cluster LP nearly optimally, using the preclustered instance $(\preclustering, \colorpreclustering, \Eadm)$ satisfying Theorem~\ref{thm:preclu:main}.
Our technique is obtained by introducing one more dimension of color to Cao et al.'s approach~\cite{cao2024understanding}.
Recall that they first formulate the \emph{bounded sub-cluster LP} making use of a constant-level Sherali-Adams hierarchy, present a procedure that samples one cluster using Raghavendra and Tan's rounding scheme~\cite{raghavendra2012approximating}, and lastly argue that the Monte Carlo method with sufficiently many samples from the procedure produces a nearly optimal solution with high probability.

In what follows, we describe the algorithm for solving the chromatic cluster LPs, adapted from Cao et al.'s approach to the chromatic versions.
We remark that there are no differences in the algorithm and analysis between \CCC and \WCCC except the objective function of the bounded sub-cluster LP.
The entire analysis is deferred to Appendix~\ref{app:sampling}.

\textbf{Bounded sub-cluster LP.}
We first present the bounded sub-cluster LP for the chromatic versions.
Given $\varepsilon > 0$ sufficiently small, we define $\varepsilon_1 := \varepsilon^3$, $\epsrt := \varepsilon^2_1 = \varepsilon^6$, and $r := \Theta(\epsrt^{-2}) = \Theta(\varepsilon^{-12})$.
We assume that $r$ is an integer where a large enough constant is hidden in its $\Theta$-notation.
For $u \in V$, let $K_u \in \preclustering$ be the precluster containing $u$.
For $u\in V$, let $\Nadm(u):=\{v\in V \mid uv \in \Eadm\}$; for $K \in \preclustering$, let $\Nadm(K) := \{v \in V \mid \exists u \in K, uv \in \Eadm\}$.
Then, the bounded sub-cluster LP is defined as follows:
\begin{align*}
    \text{min } & \obj(x) \nonumber 
    \\
    \text{s.t. }
    & \textstyle y^{\ell}_S = \sum_{s=1}^n y^{\ell, s}_S,
    && \forall \ell \in L \; \forall S \subseteq V : |S| \leq r, 
    \\
    & t^\ell_v = 1 - y^\ell_v,
    && \forall \ell \in L \; \forall v \in V,
    \\
    & x^\ell_{uv} = 1 - y^\ell_{uv}, 
    && \forall \ell \in L \; \forall uv \in \textstyle\binom{V}{2},
    \\
    & \textstyle \sum_{\ell \in L} y^\ell_v = 1,
    && \forall v \in V,
    \\
    & \textstyle \sum_{v \in V} y^{\ell, s}_{Sv} = s \; y^{\ell, s}_S, 
    && {\forall \ell \in L \; \forall s \in [n] \; \forall S \subseteq V : |S| \leq r-1,}
    \\
    & x^{\colorpreclustering(K)}_{uv} = 0, 
    && \forall uv \subseteq K \in \preclustering_{>1},
    \\
    & x^{\ell}_{uv} = 1, 
    && \forall uv \subseteq K \in \preclustering_{>1} \; \forall \ell \neq \colorpreclustering(K),
    \\
    & y^{\ell, s}_{uv} = y^{\ell, s}_u,
    && \forall \ell \in L \; \forall s \in [n] \; \forall u \in V \; \forall v \in K_u,
    \\
    & x^{\ell}_{uv} = 1,
    && \forall \text{non-admissible $uv$} \; \forall \ell \in L, 
    \\
    & y^{\ell, s}_v = 0, 
    \nonumber \\
    & \mathrlap{\hspace{5em} \forall \ell \in L \; \forall v \in V \; \forall s \in [|K_v| - 1] \cup [|K_v| + 1, |K_v| + \varepsilon_1 |\Nadm(v)|], }
    \\
    & y^{\colorpreclustering(K), |K|}_S = y^{\colorpreclustering(K), |K|}_v, 
    && \forall v \in K \in \preclustering_{>1} \; \forall S \subseteq K : |S| \leq r, 
    \\
    & y^{\colorpreclustering(K), |K|}_S = 0,
    && \forall K \in \preclustering_{>1} \; \forall S \not\subseteq K : |S| \leq r, S\cap K \neq \emptyset,
    \\
    & \textstyle\sum_{T' \subseteq T} (-1)^{|T'|} y^{\ell, s}_{S \cup T'} \in [0, y^{\ell, s}_S],
    \nonumber \\
    & \mathrlap{\hspace{5em} \forall \ell \in L \; \forall s \in [n] \; \forall S, T \subseteq V : S \cap T = \emptyset, |S \cup T| \leq r,}
    \\
    & y^{\ell, s}_S \geq 0,
    && \forall \ell \in L \; \forall s \in [n] \; \forall S \subseteq V : |S| \leq r, 
\end{align*}
where $y^{\ell, s}_S = 1$ indicates whether $S$ is a subset of a cluster of size $s$ colored $\ell$.
The only distinction between \CCC and \WCCC is in the objective function, $\obj(x)$.
Since the size of the bounded sub-cluster LP is indeed bounded by $n^{\mathsf{poly}(\varepsilon^{-1})}$, we can solve this LP in time $n^{\mathsf{poly}(\varepsilon^{-1})}$.

\textbf{Sampling one colored cluster.}
We now present the procedure for sampling one cluster from the bounded sub-cluster LP.
Let $y$ be an optimal solution to the bounded sub-cluster LP.
The procedure randomly samples a color $\ell \in L$ with probability $y^\ell_\emptyset/y_\emptyset$ where $y_\emptyset := \sum_{\ell' \in L} y^{\ell'}_\emptyset$, a cardinality $s \in [n]$ with probability $y^{\ell, s}_\emptyset/y^\ell_\emptyset$, and a pivot $u \in V$ with probability $y^{\ell, s}_u/(s \cdot y^{\ell, s}_\emptyset)$.
With these sampled $\ell$, $s$, and $u$, define $y'$ as $y'_S := y^{\ell, s}_{Su}/y^{\ell, s}_{u}$ for every $S \subseteq V$ of size $|S| \leq r-1$.
The procedure then runs Raghavendra and Tan's rounding scheme~\cite{raghavendra2012approximating} upon $y'$ to sample a cluster $C \subseteq V$, and return $C$ colored $\chi(C) := \ell$.

\textbf{Monte Carlo method.}
Finally, we describe the Monte Carlo method for constructing a nearly optimal solution.
Let $\Delta := \Theta \left( \frac{n^4 \log n}{\varepsilon_1^2 |\Eadm|} \right)$ with a sufficiently large hidden constant.
(Here, we assume $\Eadm\neq\emptyset$; otherwise, $(\preclustering,\colorpreclustering)$ given by Theorem~\ref{thm:preclu:main}, with arbitrary colors assigned to singleton preclusters, is already a $(1+\varepsilon)$-approximate integral solution.)
In particular, we assume that $\Delta y_\emptyset$ is an integer and $\Delta \geq \frac{1}{\varepsilon - \varepsilon_1} = \frac{1}{\varepsilon - \varepsilon^3}$.
The method independently samples colored clusters from the above procedure for $\Delta y_\emptyset$ times.
Let $C_1, \ldots, C_{\Delta y_\emptyset}$ be the sampled colored clusters.
Let $\ell_1, \ldots, \ell_{\Delta y_\emptyset}$ denote their corresponding colors.
For each vertex $v \in V$, let $R_v := \{t \in \{1, 2, \ldots, \Delta y_\emptyset \} \mid v \in C_t \}$, and let $\Rtilde_v$ be the set of the $\ceil{(1 - \varepsilon) \Delta}$ smallest indices in $R_v$.
For every $t \in \{1, 2, \ldots, \Delta y_\emptyset \}$, we define $\Ctilde_t := \{ v \in V \mid t \in \Rtilde_v \} \subseteq C_t$.
The final output is $\ztilde^\ell_S := \frac{|\{ t \mid \Ctilde_t = S \wedge \ell_t = \ell \}|}{\ceil{(1 - \varepsilon) \Delta}}$ for every $\ell \in L$ and $S \subseteq V$.

\bibliography{ccc}

\newpage
\appendix
\section{Deferred analysis from Section~\ref{sec_lp}} \label{app:lp}
We can rephrase the objective function for \CCC as follows:
\begin{align*}
    & 
    \textstyle \sum_{S \subseteq V, \ell \in L} \left( \frac{|\delta^+(S)|}{2} + |E^{-\ell}[S]| \right) z^\ell_S
    \\ & 
    \textstyle = \sum_{uv \in E^+} \left[ \frac{1}{2} \sum_{\ell \in L} \left( \sum_{S \ni u, S \not\ni v} z^\ell_S + \sum_{S \not\ni u, S \ni v} z^\ell_S \right) + \sum_{S \supseteq uv, \ell \neq c_{uv}} z^\ell_S \right]
    \\ &
    \textstyle \qquad + \sum_{uv \in E^-} \sum_{\ell \in L} \sum_{S \supseteq uv} z^\ell_S
    \\ &
    \textstyle = \sum_{uv \in E^+} \left[ \sum_{\ell \in L} x^\ell_{uv} - \sum_{\ell \in L}\frac{t^\ell_u + t^\ell_v}{2} + \sum_{\ell \neq c_{uv}} \sum_{S \supseteq uv} z^\ell_S \right]
    \\ &
    \textstyle \qquad  + \sum_{uv \in E^-} \sum_{\ell \in L} \sum_{S \supseteq uv} z^\ell_S
    \\ &
    \textstyle = \sum_{uv \in E^+} \left[ \sum_{\ell \in L} x^\ell_{uv} - (|L| - 1) + \sum_{\ell \neq c_{uv}} (1 - x^\ell_{uv}) \right]
    \\ &
    \textstyle \qquad + \sum_{uv \in E^-} \sum_{\ell \in L} \sum_{S \supseteq uv} z^\ell_S
    \\ &
    \textstyle = \sum_{uv \in E^+} x^{c_{uv}}_{uv} + \sum_{uv \in E^-} \sum_{\ell \in L} (1 - x^\ell_{uv})
    =: \obj(x),
\end{align*}
where the second equality comes from \eqref{eq:xminust} and the third from \eqref{eq:sumt}.

Using a similar derivation, we can also rephrase the objective function for \WCCC as follows:
\begin{align*}
    & \sum_{S \subseteq V, \ell \in L} \left( \frac{w^+(S,V\setminus S)}{2} + w^{-\ell}(S,S) \right) z^\ell_S
    \\ & 
    = \sum_{uv \in E} \left[ \frac{w^+(u,v)}{2} \sum_{\ell \in L} \left( \sum_{S \ni u, S \not\ni v} z^\ell_S + \sum_{S \not\ni u, S \ni v} z^\ell_S \right) + \sum_{S \supseteq uv}\sum_{\ell \in L} w^{-\ell}(u,v)\cdot z^\ell_S \right]
    \\ &
    = \sum_{uv \in E} \left[ w^+(u,v)\cdot \left( \sum_{\ell \in L} x^\ell_{uv} - \sum_{\ell \in L}\frac{t^\ell_u + t^\ell_v}{2}\right) + \sum_{S \supseteq uv}\sum_{\ell \in L} w^{-\ell}(u,v)\cdot z^\ell_S \right]
    \\ &
    = \sum_{uv \in E} \left[ w^+(u,v)\cdot \left( \sum_{\ell \in L} x^\ell_{uv} - (|L| - 1) \right) + \sum_{\ell \in L} w^{-\ell}(u,v)\cdot (1 - x^\ell_{uv}) \right]
    \\ &
    = \sum_{uv \in E} \left[ \sum_{\ell \in L} (w^+(u,v)-w^{-\ell}(u,v)) \cdot x^\ell_{uv} - w^+(u,v)(|L| - 1) + \sum_{\ell \in L} w^{-\ell}(u,v) \right]
    \\ &
    = \sum_{uv \in E} \left[\sum_{\ell \in L} (w^\ell(u,v)-w^{-}(u,v)) \cdot x^\ell_{uv} + w^-(u,v)|L|\right] 
    \\ &
    = \sum_{uv \in E} \sum_{\ell \in L} \left( w^\ell(u,v) x^\ell_{uv} + w^-(u,v)(1 - x^\ell_{uv}) \right)
    \\ &
    =: \obj(x),
\end{align*}
where the second equality comes from \eqref{eq:xminust}, the third from \eqref{eq:sumt}, and the fifth from the definition of $w^+$ and $w^{-\ell}$.

\begin{proof} [Proof of Lemma~\ref{lem:alg:key}]
    Consider the set $S$ such that $S \cap \{u, v\} \neq \emptyset$ for the first time.
    We have
    \begin{align*}
        \textstyle \sum_{\ell' \in L, S : S \cap uv \neq \emptyset} z^{\ell'}_S 
        &
        \textstyle = \sum_{\ell' \in L, S \ni u} z^{\ell'}_S + \sum_{\ell' \in L, S \ni v} z^{\ell'}_S - \sum_{\ell' \in L, S \supseteq uv} z^{\ell'}_S
        \\&
        = 1 + x_{uv}.
    \end{align*}
    For the first statement, recall that $\sum_{S \supseteq uv} z^\ell_S = 1 - x^\ell_{uv}$.
    The second statement immediately follows from the fact that $\sum_{\ell' \in L, S \supseteq uv} z^{\ell'}_S = 1 - x_{uv}$.
\end{proof}

\section{Deferred analysis from Section~\ref{sec:preclu}}
\subsection{Constructing \texorpdfstring{$\preclustering$}{K} and \texorpdfstring{$\colorpreclustering$}{chi pre} for \CCC} \label{app:preclu:K}

Recall that we use parameters $\alpha,\beta\in(0,1)$ such that $\eta:=\frac{\alpha+\beta}{1-\beta}\in (0,\frac{1}{13})$.

\begin{lemma}[Restatement of Lemma~\ref{lem-main:precluster-disagreement-bound}]\label{lem:precluster-disagreement-bound}
    For any non-singleton precluster $K \in \preclustering_{>1}$ with $\ell:=\colorpreclustering(K)$ and $u \in K$, we have
    \[
        |E^{-\ell}(u,K)| < \eta (|K|-1) 
        \text{ and }
        |E^+(u,V\setminus K)|< \eta (|K|-1).
    \]
\end{lemma}
\begin{proof}
Let $C\in\initialclustering$ be the cluster containing $K\subseteq C$. 
Since $K$ is non-singleton, the vertices in $K$ remain unmarked during the execution of the algorithm, implying that
\begin{equation} \label{eq:preclu:dab01}
    |C\setminus K| <\beta (|C|-1).
\end{equation}
Since $|C \setminus K| = (|C| - 1) - (|K| - 1)$, rearranging the terms yields 
\begin{equation} \label{eq:preclu:dab02}
    |C|-1<\frac{1}{1-\beta}(|K|-1).
\end{equation}
Let $\ell := \colorpreclustering(K) = \colorinitialclustering(C)$.
We can also observe that
\begin{equation} \label{eq:preclu:dab03}
    |E^{-\ell}(u,C)| < \alpha (|C|-1) \text{ and } |E^+(u,V\setminus C)|< \alpha (|C|-1)
\end{equation}
since $u \in K$ is unmarked.
We therefore have
\begin{align*}
    |E^{-\ell}(u,K)|\le|E^{-\ell}(u,C)|<\alpha (|C|-1) < \frac{\alpha}{1-\beta}(|K|-1)<\eta (|K|-1),
\end{align*}
where the first inequality follows from the fact that $E^{-\ell}(u,K) \subseteq E^{-\ell}(u,C)$ and the second-to-last inequality is due to Inequality~\eqref{eq:preclu:dab02}.
Furthermore, we have
\begin{align*}
    |E^+(u,V\setminus K)| 
    & 
    = |E^+(u,C\setminus K)|+|E^+(u,V\setminus C)|
    \leq |C \setminus K| + |E^+(u,V\setminus C)|
    \\ &
    < \beta (|C| - 1) + \alpha (|C| - 1)
    < \frac{\alpha + \beta}{1 - \beta} (|K| - 1)
    = \eta \, (|K| - 1),
\end{align*}
where the second inequality is derived from Inequalities~\eqref{eq:preclu:dab01} and~\eqref{eq:preclu:dab03} and the last from Inequality~\eqref{eq:preclu:dab02}.
\end{proof}

In what follows, we may instead use the following slightly weaker inequalities in most cases:
\[
    |E^{-\ell}(u, K)| < \eta |K| \text{ and } |E^+(u, V \setminus K)| < \eta |K|.
\]

Now, fix any optimal solution $(\scriptCstar,\chistar)$ to the given instance $I$.

\begin{lemma}[Restatement of Lemma~\ref{lem-main:precluster-domination} for \CCC]\label{lem:precluster-domination}
    For any non-singleton precluster $K\in\preclustering_{>1}$, if a cluster $\Cstar_1\in\scriptCstar$ colored $\colorpreclustering(K)$ intersects $K$ (i.e., $\chistar(\Cstar_1) = \colorpreclustering(K)$ and $K \cap \Cstar_1 \neq \emptyset$), we have $|\Cstar_1 \setminus K| < 2\eta |K|$.
\end{lemma}
\begin{proof}
    Let $\ell:=\colorpreclustering(K)=\chistar(\Cstar_1)$ and $K_1:=K\cap \Cstar_1$.
    We define a solution $(\scriptC,\chi)$ constructed from $(\scriptCstar, \chistar)$ by splitting $\Cstar_1$ into $K_1$ and $\Cstar_1 \setminus K_1$.
    That is,
    \begin{equation*}
        \scriptC := \scriptCstar \setminus \{\Cstar_1\} \cup \{ K_1, \Cstar_1 \setminus K_1 \} 
        \text{ and }
        \chi(C) := \begin{cases}
            \ell, & \text{if $C \in \{K_1, \Cstar_1 \setminus K_1 \}$,} \\
            \chistar(C), & \text{if $C \in \scriptC \cap \scriptCstar$.}
        \end{cases}
    \end{equation*}
    Suppose toward contradiction that $|\Cstar_1 \setminus K| \ge 2\eta |K|$. We show that the number of disagreements in $(\scriptC,\chi)$ is less than that of $(\scriptCstar,\chistar)$. To this end, we classify all edges as follows, and determine whether each edge is a disagreement or not in $(\scriptC,\chi)$ and in $(\scriptCstar,\chistar)$:
    \begin{enumerate}
        \item Each $-$ edge across $K_1$ and $\Cstar_1 \setminus K_1$ is an agreement in $(\scriptC,\chi)$, but a disagreement in $(\scriptCstar,\chistar)$.\label{case0:gain}
        \item Each $-$ edge across $K_1$ and $V \setminus \Cstar_1$ is an agreement in both solutions. 
        \item Each $+$ edge colored $\ell$ across $K_1$ and $\Cstar_1 \setminus K_1$ is a disagreement in $(\scriptC,\chi)$, but an agreement in $(\scriptCstar,\chistar)$.\label{case0:loss}
        \item Each $+$ edge colored differently from $\ell$ across $K_1$ and $\Cstar_1 \setminus K_1$ is a disagreement in both solutions.
        \item Each $+$ edge across $K_1$ and $V \setminus \Cstar_1$ is a disagreement in both solutions.
        \item Each edge within $K_1$ has the same status in both solutions.
        \item The remaining edges outside $\Cstar_1$ also have the same statuses in both solutions, respectively.
    \end{enumerate}
    
    We first lower bound the ``gain'' in cost of the constructed solution $(\scriptC, \chi)$ --- the number of edges that are agreed in $(\scriptC,\chi)$, but not in $(\scriptCstar, \chistar)$. 
    Observe that the type~\ref{case0:gain} edges contribute to this gain.
    For any $u \in K_1$, note that
    \begin{align*}
        |E^{-}(u,\Cstar_1\setminus K_1)| 
        &
        = |E(u,\Cstar_1\setminus K_1)| - |E^{+}(u,\Cstar_1\setminus K_1)|
        \\&
        = |E(u,\Cstar_1\setminus K_1)| - |E^{+}(u,\Cstar_1\setminus K)|
        \\&
        \ge |E(u,\Cstar_1\setminus K_1)| - |E^{+}(u,V\setminus K)| 
        \\&
        =|\Cstar_1\setminus K_1| - |E^{+}(u,V\setminus K)| 
        \\&
        > |\Cstar_1\setminus K_1| - \eta |K|,
    \end{align*}
    where the last equality comes from $u \not\in \Cstar_1 \setminus K_1$, and the last inequality from Lemma~\ref{lem:precluster-disagreement-bound} (recall that $K$ is non-singleton).
    We can thus derive that the gain of $(\scriptC, \chi)$, the total number of type~\ref{case0:gain} edges, is strictly greater than $|K_1| \cdot (|\Cstar_1\setminus K_1| - \eta |K|) = |K_1| \cdot (|\Cstar_1 \setminus K| - \eta |K|)$.

    We now upper bound the ``loss'' in cost of $(\scriptC, \chi)$ --- the number of edges that are agreed in $(\scriptCstar, \chistar)$, but not in $(\scriptC, \chi)$.
    Note that the edges of type~\ref{case0:loss} contribute to this loss.
    For any $u \in K_1$, we have
    \begin{align*}
        |E^{\ell}(u,\Cstar_1\setminus K_1)| = |E^{\ell}(u,\Cstar_1\setminus K)| \le |E^{+}(u,V\setminus K)|< \eta |K|
    \end{align*}
    where the last inequality is again due to Lemma~\ref{lem:precluster-disagreement-bound}, yielding that the loss of $(\scriptC, \chi)$ is strictly less than $|K_1| \cdot \eta |K|$.
    
    We can thus conclude that the ``net gain'' in cost of the constructed solution $(\scriptC, \chi)$ against $(\scriptCstar, \chistar)$ is strictly greater than
    \[
        |K_1|\cdot (|\Cstar_1\setminus K| - \eta |K|) - |K_1|\cdot \eta |K| 
        = |K_1| \cdot (|\Cstar_1\setminus K| - 2 \eta |K|)
        \ge 0,
    \]
    where the inequality is due to the assumption that $|\Cstar_1 \setminus K| \ge 2\eta |K|$, contradicting the optimality of $(\scriptCstar, \chistar)$.
\end{proof}

\begin{lemma}[Restatement of Lemma~\ref{lem-main:precluster-subdivide}]\label{lem:precluster-subdivide}
    $\preclustering$ subdivides $\scriptCstar$, i.e., for any $K\in\preclustering$, there is $\Cstar \in \scriptCstar$ such that $K \subseteq \Cstar$.
\end{lemma}
\begin{proof}
    Suppose toward contradiction that $\scriptCstar$ is not subdivided by $\preclustering$.
    Then, there exists a non-singleton precluster $K \in \preclustering_{>1}$ such that $K \setminus \Cstar$ is nonempty for every cluster $\Cstar \in \scriptCstar$.
    We choose an arbitrary such precluster $K \in \preclustering_{>1}$, and let $\ell:=\colorpreclustering(K)$ be the color of $K$ in the preclustering.
    Let $\xi$ be a constant between $\frac{1+5\eta}{2}$ and $1-4\eta$.
    Note that we can always choose such $\xi$ because $\frac{1+5\eta}{2}< 1-4\eta$ for $\eta \in (0,\frac{1}{13})$.
    Observe that $\frac{1}{2}<\xi< 1$.

    We break the analysis into two cases depending on the existence of a cluster $\Cstar \in \scriptCstar$ that ``dominantly'' intersects $K$, i.e., $|K \cap \Cstar| \geq \xi |K|$.
    The overall structure of the proof for each case is similar to the proof of Lemma~\ref{lem:precluster-domination}: we respectively construct a solution whose cost is strictly less than $(\scriptCstar, \chistar)$, leading to contradiction that $(\scriptCstar, \chistar)$ is optimal.
    
    \textbf{Case 1. No dominant intersections.} Suppose that, for all $\Cstar \in \scriptCstar$, $|K \cap \Cstar| < \xi |K|$.
    Consider a new solution $(\scriptC',\chi')$ constructed by inserting $K$ colored $\ell = \colorpreclustering(K)$. More precisely, we have
    \begin{align*}
        \scriptC' & := \{ K \} \cup \{\Cstar \setminus K \mid \Cstar \in \scriptCstar \text{ with } |\Cstar \setminus K| > 0\} \text{ and }
        \\
        \chi'(C) & := \begin{cases}
            \ell, & \text{if $C = K$}; \\
            \chistar(\Cstar), & \text{if $C = \Cstar \setminus K$.}
        \end{cases}
    \end{align*}
    We classify each edge and determine whether it is a disagreement or not in both solutions.
    \begin{enumerate}
        \item Each $-$ edge across $K$ and $V\setminus K$ is agreed in $(\scriptC', \chi')$, and possibly agreed in $(\scriptCstar, \chistar)$.
        \item Each $+$ edge across $K$ and $V\setminus K$ is disagreed in $(\scriptC', \chi')$, but possibly agreed in $(\scriptCstar, \chistar)$. \label{case1:loss1}
        \item Each $-$ edge within $K$ is disagreed in $(\scriptC', \chi')$, but possibly agreed in $(\scriptCstar, \chistar)$. \label{case1:loss2}
        \item Each $+$ edge $e$ of color $c_e = \ell$ within $K\cap \Cstar$ for some $\Cstar\in\scriptCstar$ is agreed in $(\scriptC', \chi')$, and possibly agreed in $(\scriptCstar, \chistar)$.
        \item Each $+$ edge $e$ of color $c_e \neq \ell$ within $K\cap \Cstar$ for some $\Cstar\in\scriptCstar$ is disagreed in $(\scriptC', \chi')$, but possibly agreed in $(\scriptCstar, \chistar)$. \label{case1:loss3}
        \item Each $+$ edge $e$ of color $c_e=\ell$ that is within $K$ but across clusters in $\scriptCstar$ is agreed in $(\scriptC', \chi')$, but disagreed in $(\scriptCstar, \chistar)$. \label{case1:gain}
        \item Each $+$ edge $e$ of color $c_e\neq\ell$ that is within $K$ but across clusters in $\scriptCstar$ is disagreed in both solutions.
        \item The remaining edges outside $K$ have the same statuses in both solutions, respectively.
    \end{enumerate}
    Observe that the gain in cost of $(\scriptC', \chi')$ against $(\scriptCstar, \chistar)$ is lower-bounded by the number of type~\ref{case1:gain} edges while the loss in cost of $(\scriptC', \chi')$ is upper-bounded by the number of edges of types~\ref{case1:loss1}, \ref{case1:loss2}, and \ref{case1:loss3}.
    
    To give a lower bound on the number of type~\ref{case1:gain} edges, we consider any $u \in K$ and let $\Cstar_u \in \scriptCstar$ be the cluster containing $u$ in $\scriptCstar$. 
    We then have
    \begin{align*}
        |E^{\ell}(u,K\setminus \Cstar_u)|
        &
        = |E(u,K\setminus \Cstar_u)| - |E^{-\ell}(u,K \setminus \Cstar_u)|
        \\&
        \ge |E(u,K\setminus \Cstar_u)| - |E^{-\ell}(u,K)| 
        \\&
        =|K\setminus \Cstar_u|- |E^{-\ell}(u,K)| 
        > |K\setminus \Cstar_u|- \eta |K|
        \\&
        \ge (1-\xi)|K|- \eta |K| = (1-\xi-\eta)|K|,
    \end{align*}
    where the second equality comes from $u \not\in K \setminus \Cstar_u$, the second inequality from Lemma~\ref{lem:precluster-disagreement-bound}, and the last inequality from the condition of this case. 
    Hence, the number of type~\ref{case1:gain} edges is bounded from below by
    \[
        \frac{1}{2} \sum_{u \in K} |E^\ell (u, K \setminus \Cstar_u)| > \frac{1 - \xi - \eta}{2} |K|^2.
    \]
    This is also a lower bound for the gain in cost of $(\scriptC', \chi')$ against $(\scriptCstar, \chistar)$.
    
    We now consider the edges of types~\ref{case1:loss1}, \ref{case1:loss2} and \ref{case1:loss3}.
    Due to the second inequality of Lemma~\ref{lem:precluster-disagreement-bound}, we can upper bound the number of type~\ref{case1:loss1} edges by $|K| \cdot \eta |K| = \eta |K|^2$.
    On the other hand, the total number of type~\ref{case1:loss2} and type~\ref{case1:loss3} edges is bounded from above by
    \[
        \frac{1}{2} \sum_{u \in K} |E^{-\ell}(u, K)| < \frac{\eta}{2} |K|^2
    \]
    due to the first inequality of Lemma~\ref{lem:precluster-disagreement-bound}.
    Hence, the loss in cost of $(\scriptC', \chi')$ against $(\scriptCstar, \chistar)$ is strictly less than $\frac{3\eta}{2} |K|^2$.
    Since $\xi\le 1-4\eta$, we have
    \[
        \frac{1-\xi-\eta}{2} |K|^2 - \frac{3\eta}{2}|K|^2\ge 0,
    \]
    implying that $(\scriptC',\chi')$ has fewer disagreements than $(\scriptCstar,\chistar)$.
    This contradicts the optimality of $(\scriptCstar, \chistar)$.

    \textbf{Case 2. A dominant intersection exists.}
    Suppose now there exists a cluster $\Cstar_1 \in \scriptCstar$ such that $|K \cap \Cstar_1| \ge \xi |K|$.
    Note that we always have at most one such cluster since $\xi>\frac{1}{2}$.
    Let $K_1:=K \cap \Cstar_1$.
    
    We first claim that $\chistar(\Cstar_1) \neq \colorpreclustering(K)$.
    Otherwise, if $\chistar(\Cstar_1) = \colorpreclustering(K)$, we construct a solution $(\scriptC'',\chi'')$ by transferring one arbitrary vertex $w \in K \setminus \Cstar_1$ to $\Cstar_1$.
    In other words,
    \begin{align*}
        \scriptC'' & := \{ \Cstar_1 \cup \{w\} \} \cup \{ \Cstar \setminus \{w\} \mid \Cstar \in \scriptCstar \setminus \{\Cstar_1\} \} \text{ and}
        \\
        \chi''(C) & = \begin{cases}
            \chistar(\Cstar_1), & \text{if $C = \Cstar_1 \cup \{w\}$,} \\
            \chistar(\Cstar), & \text{if $C = \Cstar \setminus \{w\}$.}
        \end{cases}
    \end{align*}
    Since $K \setminus \Cstar_1$ is nonempty, we can always choose $w \in K \setminus \Cstar_1$ in the above construction.
    Recall also that $\ell = \colorpreclustering(K) = \chistar(\Cstar_1)$. Consider the following classification of edges:
    \begin{enumerate}
        \item Each edge in $E^{\ell}(w,\Cstar_1)$ is agreed in $(\scriptC'', \chi'')$, but disagreed in $(\scriptCstar, \chistar)$. \label{case2-1:gain}
        \item Each edge in $E^{-\ell}(w,\Cstar_1)$ is disagreed in $(\scriptC'', \chi'')$, but possibly agreed in $(\scriptCstar, \chistar)$. \label{case2-1:loss1}
        \item Each edge in $E^{+}(w,V\setminus \Cstar_1)$ is disagreed in $(\scriptC'', \chi'')$, but possibly agreed in $(\scriptCstar, \chistar)$. \label{case2-1:loss2}
        \item Each edge in $E^{-}(w,V\setminus \Cstar_1)$ is agreed in $(\scriptC'', \chi'')$, and possibly agreed in $(\scriptCstar, \chistar)$.
        \item All the edges that are not incident with $w$ have the same statuses in both solutions, respectively.
    \end{enumerate}
    Hence, the gain in cost of $(\scriptC'', \chi'')$ against $(\scriptCstar, \chistar)$ is bounded from below by $|E^\ell(w, \Cstar_1)|$, whilst the loss in cost of $(\scriptC'', \chi'')$ is bounded from above by $|E^{-\ell}(w, \Cstar_1)| + |E^+(w, V \setminus \Cstar_1)|$. 
    For the type~\ref{case2-1:gain} edges, we have
    \begin{align*}
        |E^{\ell}(w,\Cstar_1)|
        &
        \ge |E^{\ell}(w,K_1)|
        =|E(w,K_1)|-|E^{-\ell}(w,K_1)|
        \\&
        =|K_1|-|E^{-\ell}(w,K_1)|
        \ge |K_1|-|E^{-\ell}(w,K)|
        \\&
        > |K_1|-\eta |K|
        \ge \xi|K|-\eta |K| = (\xi - \eta) |K|,
    \end{align*}
    where the second equality comes from $w \notin K_1$, and the second-to-last inequality from Lemma~\ref{lem:precluster-disagreement-bound}.
    For type~\ref{case2-1:loss1} edges,
    \begin{align*}
        |E^{-\ell}(w,\Cstar_1)| 
        &
        =|E^{-\ell}(w,K_1)|+|E^{-\ell}(w,\Cstar_1\setminus K_1)| 
        \\&
        \le |E^{-\ell}(w,K)| + |\Cstar_1\setminus K_1|
        \\&
        = |E^{-\ell}(w,K)| + |\Cstar_1\setminus K|
        \\&
        < \eta |K| + 2\eta |K| = 3\eta |K|,
    \end{align*}
    where the last inequality comes from Lemma~\ref{lem:precluster-disagreement-bound} and Lemma~\ref{lem:precluster-domination}.
    Lastly, for type~\ref{case2-1:loss2} edges,
    \begin{align*}
        |E^{+}(w,V\setminus \Cstar_1)| 
        & 
        = |E^{+}(w,K \setminus \Cstar_1)|+|E^{+}(w,(V\setminus K)\setminus \Cstar_1)|
        \\&
        < |K \setminus \Cstar_1|+|E^{+}(w,V\setminus K)|
        \\&
        \le (1-\xi)|K|+|E^{+}(w,V\setminus K)|
        \\& 
        < (1-\xi)|K|+\eta |K| = (1 - \xi + \eta) |K|,
    \end{align*}
    where the second inequality comes from that $|K \cap \Cstar_1| \geq \xi |K|$, and the last inequality is due to Lemma~\ref{lem:precluster-disagreement-bound}.
    Hence, the net gain in cost of $(\scriptC'', \chi'')$ against $(\scriptCstar, \chistar)$ is strictly greater than
    \[
        (\xi - \eta) |K| - (1 - \xi + 4\eta) |K| \geq 0,
    \]
    where the inequality follows from the choice of $\xi > \frac{1 + 5\eta}{2}$, contradicting the optimality of $(\scriptCstar, \chistar)$.
    This completes the proof of our claim that $\chistar(\Cstar_1) \neq \colorpreclustering(K)$.

    We now construct another solution $(\scriptC, \chi)$ from $(\scriptCstar, \chistar)$ by splitting $\Cstar_1$ into $K_1$ colored $\ell = \colorpreclustering(K)$ and $\Cstar_1 \setminus K_1$ colored $\chistar(\Cstar_1)$, i.e.,
    \[
        \scriptC := \scriptCstar \setminus \{\Cstar_1\} \cup \{ K_1, \Cstar_1 \setminus K_1 \} 
        \quad\text{ and }\quad
        \chi(C) := \begin{cases}
            \ell, & \text{if $C = K_1$,} \\
            \chistar(\Cstar_1), & \text{if $C = \Cstar_1 \setminus K_1$,} \\
            \chistar(C), & \text{if $C \in \scriptC \cap \scriptCstar$.}
        \end{cases}.
    \]
    Consider the following breakdown of the edges:
    \begin{enumerate}
        \item Each $-$ edge across $K_1$ and $V \setminus K_1$ is agreed in $(\scriptC, \chi)$, and possibly agreed in $(\scriptCstar, \chistar)$.
        \item Each $+$ edge across $K_1$ and $\Cstar_1 \setminus K_1$ is disagreed in $(\scriptC,\chi)$, but possibly agreed in $(\scriptCstar,\chistar)$. \label{case2-2:loss1}
        \item Each $+$ edge across $K_1$ and $V \setminus \Cstar_1$ is disagreed in both solutions.
        \item Each $-$ edge within $K_1$ is disagreed in both solutions.
        \item Each $+$ edge $e$ of color $c_e=\ell$ within $K_1$ is agreed in $(\scriptC, \chi)$, but disagreed in $(\scriptCstar, \chistar)$. \label{case2-2:gain}
        \item Each $+$ edge $e$ of color $c_e \neq \ell$ within $K_1$ is disagreed in $(\scriptC,\chi)$, but possibly agreed in $(\scriptCstar,\chistar)$. \label{case2-2:loss2}
        \item All the remaining edges outside $K_1$ have the same statuses in both solutions, respectively.
    \end{enumerate}
    We can thus see that the gain in cost of $(\scriptC, \chi)$ against $(\scriptCstar, \chistar)$ is at least the number of type~\ref{case2-2:gain} edges, while the loss is at most the total number of type~\ref{case2-2:loss1} and type~\ref{case2-2:loss2} edges.
    
    To lower-bound the number of type~\ref{case2-2:gain} edges, we first fix an $u \in K_1$ and consider the type~\ref{case2-2:gain} edges incident to $u$. We then have
    \begin{align*}
        |E^\ell(u,K_1)| 
        &
        = |E(u,K_1)| - |E^{-\ell}(u,K_1)| 
        = |K_1|-1 - |E^{-\ell}(u,K_1)| 
        \\&
        \ge |K_1|-1 - |E^{-\ell}(u,K)| 
        > |K_1|-1 - \eta (|K|-1) 
        \\&
        \ge \xi |K| -1  - \eta (|K|-1),
    \end{align*}
    where the second inequality comes from Lemma~\ref{lem:precluster-disagreement-bound}, and the last inequality comes from the condition of $K_1$.
    Hence, the number of type~\ref{case2-2:gain} edges is at least
    \[
        \frac{1}{2} \sum_{u \in K_1} |E^\ell(u, K_1)|
        > \frac{1}{2}|K_1|\cdot(\xi|K|-1 -\eta(|K|-1)).
    \]

    We now upper-bound the number of edges of types~\ref{case2-2:loss1} and~\ref{case2-2:loss2}.
    For any $u \in K_1$, we have
    \begin{align*}
        |E^+(u,\Cstar_1 \setminus K_1)| 
        &
        = |E^+(u,\Cstar_1 \setminus K)| 
        \le |E^+(u,V \setminus K)| 
        < \eta (|K|-1),
    \end{align*}
    where the last inequality is due to Lemma~\ref{lem:precluster-disagreement-bound}, implying that the number of type~\ref{case2-2:loss1} edges is strictly less than $|K_1|\cdot \eta (|K|-1)$.
    On the other hand, we also have that, for every $u \in K_1$,
    \begin{align*}
        |E^+ \cap E^{-\ell}(u,K_1)| &\le |E^{-\ell}(u,K)|
        <\eta (|K|-1)
    \end{align*}
    where the last inequality comes from Lemma~\ref{lem:precluster-disagreement-bound}, yielding that the number of type~\ref{case2-2:loss2} edges is strictly upper-bounded by $\frac{1}{2}|K_1|\cdot \eta (|K|-1)$.

    We can thus derive that the net gain in cost of $(\scriptC, \chi)$ against $(\scriptCstar, \chistar)$ is greater than
    \begin{align*}
        & 
        \frac{1}{2}|K_1|\cdot(\xi|K|-1 -\eta(|K|-1)) - \frac{1}{2}|K_1|\cdot 3\eta(|K|-1) 
        \\&
        = \frac{1}{2}|K_1|\cdot(\xi|K|-1 -4\eta(|K|-1)).
    \end{align*}
    Observe that $\xi > \frac{1+5\eta}{2} > \frac{1}{2} > \frac{4}{13} > 4\eta$ by the choice of $\xi$ and $\eta$.
    We can thus see
    \[
        \xi|K|-1 -4\eta(|K|-1)= (\xi-4\eta)|K|-1+4\eta \ge 2\xi-4\eta-1 > \eta,
    \]
    where the first inequality comes from the fact that $K$ is non-singleton and $\xi > 4\eta$, and the last inequality from the fact that $\xi > \frac{1 + 5 \eta}{2}$.
    This again contradicts the optimality of $(\scriptCstar, \chistar)$, completing the entire proof of this case.
\end{proof}

The above lemma shows that each precluster $K \in \preclustering$ is fully contained in some cluster $\Cstar \in \scriptCstar$. 
We can also show that every non-singleton precluster must be colored the same as the cluster in $\scriptCstar$ containing it.

\begin{lemma}[Restatement of Lemma~\ref{lem-main:precluster-unique} for \CCC]\label{cor:precluster-unique}
    For every cluster $\Cstar \in \scriptCstar$ and any non-singleton precluster $K \in \preclustering$ contained in $\Cstar$, we have $\colorpreclustering(K) =  \chistar(\Cstar)$.
\end{lemma}
\begin{proof}
    Suppose toward contradiction that there exists a non-singleton $K$ and $\Cstar_1$ such that $K \subseteq \Cstar_1$ but $\colorpreclustering(K) \neq \chistar(\Cstar_1)$.
    We show that a solution $(\scriptC, \chi)$ obtained by splitting $\Cstar_1$ into $K$ of color $\colorpreclustering(K)$ and $\Cstar_1 \setminus K$ of color $\chistar(\Cstar_1)$ has fewer disagreements than $(\scriptCstar, \chistar)$.
    The proof indeed follows from exactly the same argument as in the latter half of Case 2 in the proof of Lemma~\ref{lem:precluster-subdivide}.
\end{proof}

\begin{lemma}\label{lem:precluster-constant-approx}
    $(\preclustering, \colorpreclustering)$ is a constant-approximate solution.
\end{lemma}
\begin{proof}
    We show that the number of disagreements in $(\preclustering, \colorpreclustering)$ is within a constant factor of that in $(\initialclustering,\colorinitialclustering)$; since $(\initialclustering,\colorinitialclustering)$ is a constant-approximate solution, this completes the proof.
    To this end, we charge the edges newly disagreed in $(\preclustering, \colorpreclustering)$ due to the algorithm to the edges already disagreed in $(\initialclustering, \colorinitialclustering)$.
    
    Consider any non-singleton cluster $C \in \initialclustering$ of color $\ell:=\colorinitialclustering(C)$. 
    Observe that, if a vertex $u\in C$ becomes a singleton precluster in $\preclustering$, the set of newly disagreed edges is precisely $E^\ell(u,C)$.  
    Let
    \begin{align*}
        B_1 & := \{ u \in C \mid |E^{-\ell}(u,C)| \ge \alpha (|C|-1)\} \text{ and}
        \\
        B_2 & := \{ u \in C \mid |E^{+}(u,V\setminus C)| \ge \alpha (|C|-1)\}.
    \end{align*}
    Note that $B_1 \cup B_2$ is the set of marked vertices in $C$ right after iterating over $V$.
    We break the analysis into two cases depending on the size of $B_1 \cup B_2$.
    
    \textbf{Case 1. $|B_1\cup B_2| < \beta(|C|-1)$.} 
    In this case, the vertices in $C \setminus (B_1 \cup B_2)$ remain unmarked, and hence, it suffices to consider the vertices in $B_1 \cup B_2$.
    For any $u \in B_1$, observe that
    \begin{align*}
        |E^\ell(u,C)|
        &
        = |E(u,C)|-|E^{-\ell}(u,C)|
        =|C|-1-|E^{-\ell}(u,C)|
        \\&
        \le (1 - \alpha) (|C|-1)
        \le \frac{1-\alpha}{\alpha}|E^{-\ell}(u,C)|
        <\frac{1}{\alpha}|E^{-\ell}(u,C)|,
    \end{align*}
    where the first and second inequalities are due to the definition of $B_1$; on the other hand, for $u \in B_2$, we have
    \begin{align*}
        |E^\ell(u,C)| \le |C|-1 \le \frac{1}{\alpha}|E^{+}(u,V\setminus C)|,
    \end{align*}
    where the last inequality is due to the definition of $B_2$.
    Hence, for any $u \in B_1 \cup B_2$, it is sufficient to cover $|E^\ell(u, C)|$ by charging $\frac{1}{\alpha}$ to each edge in $E^{-\ell}(u, C) \cup E^+(u, V \setminus C)$.
    Hence the total charged amount for the cluster $C$ is at most
    \[
        \frac{1}{\alpha} \left( 2 \, |\{uv \in E^{-\ell} \mid uv \subseteq C\}| + |\{uv \in E^{+} \mid |uv \cap C| = 1\}|\right).
    \]
    
    \textbf{Case 2. $|B_1\cup B_2| \ge \beta(|C|-1)$.} 
    Recall that, in this case, the algorithm marks all vertices in $C$. 
    The total number of newly disagreed edges is at most $\binom{|C|}{2}$. 
    Note that, by the definition of $B_1$,
    \[
        |\{uv \in E^{-\ell} \mid uv \subseteq C\}| 
        \ge \frac{1}{2}\sum_{u \in B_1}|E^{-\ell}(u,C)| 
        \ge \frac{1}{2} \, |B_1| \cdot \alpha (|C|-1),
    \]
    and, by the definition of $B_2$, 
    \[
        |\{uv \in E^{+} \mid |uv \cap C| = 1\}|
        \ge \sum_{u \in B_2}|E^{+}(u,V\setminus C)|
        \ge |B_2|\cdot \alpha (|C|-1).
    \]
    We can thus derive that
    \begin{align*}
        &
        2 \, |\{ uv \in E^{-\ell} \mid uv \subseteq C\}| + |\{ uv \in E^{+} \mid | uv \cap C | = 1\}| 
        \\&
        \geq \alpha (|B_1| + |B_2|) (|C| - 1)
        \\&
        \ge \alpha \beta (|C|-1)^2,
    \end{align*}
    where the last inequality is due to the condition that $|B_1| + |B_2| \geq |B_1 \cup B_2| \geq \beta (|C| - 1)$.
    This yields
    \begin{align*}
        \binom{|C|}{2}
        & 
        =\frac{|C|(|C|-1)}{2} 
        \leq (|C|-1)^2
        \\&
        \le \frac{1}{\alpha\beta} \left( 2 \, |\{ uv \in E^{-\ell} \mid uv \subseteq C\}| + |\{ uv \in E^{+} \mid | uv \cap C | = 1\}|  \right),
    \end{align*}
    where the first inequality is because $C$ is non-singleton.
    It therefore suffices to cover the new disagreements due to $C$ by charging $\frac{2}{\alpha\beta}$ to each edge in $\{uv \in E^{-\ell} \mid uv \subseteq C\}$ and $\frac{1}{\alpha\beta}$ to each edge in $\{ uv \in E^{+} \mid |uv \cap C| = 1\}$, respectively.

    Due to the above charging argument, we can see that every disagreement in $(\initialclustering, \colorinitialclustering)$ is charged at most $\frac{2}{\alpha \beta}$ while sufficiently covering the cost incurred by the new disagreements in $(\preclustering, \colorpreclustering)$.
\end{proof}

\subsection{Constructing \texorpdfstring{$\preclustering$}{K} and \texorpdfstring{$\colorpreclustering$}{chi pre} for \WCCC} \label{app:preclu:K_weighted}
Recall that we use parameters $\alpha,\beta\in(0,1)$ such that $\eta:=\frac{\alpha+\beta}{1-\beta}\in (0,\frac{1}{13})$.

\begin{lemma}[Restatement of Lemma~\ref{lem-main:precluster-disagreement-bound-weighted}]\label{lem:precluster-disagreement-bound_weighted}
    For any non-singleton precluster $K \in \preclustering_{>1}$ with $\ell:=\colorpreclustering(K)$ and $u \in K$, we have
    \[
    w^{-\ell}(u,K) < \eta (|K|-1) \text{ and } w^+(u,V\setminus K)< \eta (|K|-1).
    \]
\end{lemma}
\begin{proof}
Let $C\in\initialclustering$ be the cluster containing $K\subseteq C$. 
Since $K$ is non-singleton, the vertices in $K$ remain unmarked during the execution of the algorithm, implying that
\begin{equation} \label{eq:preclu:dab01_weighted}
    |C\setminus K| <\beta (|C|-1).
\end{equation}
Since $|C \setminus K| = (|C| - 1) - (|K| - 1)$, rearranging the terms yields 
\begin{equation} \label{eq:preclu:dab02_weighted}
    |C|-1<\frac{1}{1-\beta}(|K|-1).
\end{equation}
Let $\ell := \colorpreclustering(K) = \colorinitialclustering(C)$.
We can also observe that
\begin{equation} \label{eq:preclu:dab03_weighted}
    w^{-\ell}(u,C) < \alpha (|C|-1) \;\text{ and }\; w^+(u,V\setminus C)< \alpha (|C|-1)
\end{equation}
since $u \in K$ is unmarked.
We therefore have
\begin{align*}
    w^{-\ell}(u,K)\le w^{-\ell}(u,C)<\alpha (|C|-1) < \frac{\alpha}{1-\beta}(|K|-1)<\eta (|K|-1),
\end{align*}
where the first inequality follows from $K \subseteq C$, and the second-to-last inequality is due to Inequality~\eqref{eq:preclu:dab02_weighted}.
Furthermore, we have
\begin{align*}
    w^+(u,V\setminus K)
    & 
    = w^+(u,C\setminus K)+w^+(u,V\setminus C)
    \leq |C \setminus K| + w^+(u,V\setminus C)
    \\ &
    < \beta (|C| - 1) + \alpha (|C| - 1)
    < \frac{\alpha + \beta}{1 - \beta} (|K| - 1)
    = \eta \, (|K| - 1),
\end{align*}
where the first inequality is from the probability constraints, the second inequality is derived from Inequalities~\eqref{eq:preclu:dab01_weighted} and~\eqref{eq:preclu:dab03_weighted} and the last from Inequality~\eqref{eq:preclu:dab02_weighted}.
\end{proof}

In what follows, we may instead use the following slightly weaker inequalities in most cases:
\[
    w^{-\ell}(u, K) < \eta |K| \;\text{ and }\; w^+(u, V \setminus K) < \eta |K|.
\]

Now, fix any optimal solution $(\scriptCstar,\chistar)$ to the given instance.

\begin{lemma}[Restatement of Lemma~\ref{lem-main:precluster-domination} for \WCCC]\label{lem:precluster-domination_weighted}
    For any non-singleton precluster $K\in\preclustering_{>1}$, if a cluster $\Cstar_1\in\scriptCstar$ colored $\colorpreclustering(K)$ intersects $K$ (i.e., $\chistar(\Cstar_1) = \colorpreclustering(K)$ and $K \cap \Cstar_1 \neq \emptyset$), we have $|\Cstar_1 \setminus K| < 2\eta |K|$.
\end{lemma}
\begin{proof}
    Suppose toward contradiction that $|\Cstar_1 \setminus K| \ge 2\eta |K|$.
    Let $\ell:=\colorpreclustering(K)=\chistar(\Cstar_1)$ and $K_1:=K\cap \Cstar_1$.
    We define a solution $(\scriptC,\chi)$ constructed from $(\scriptCstar, \chistar)$ by splitting $\Cstar_1$ into $K_1$ and $\Cstar_1 \setminus K_1$.
    That is,
    \begin{align*}
        \scriptC 
        & 
        := \scriptCstar \setminus \{\Cstar_1\} \cup \{ K_1, \Cstar_1 \setminus K_1 \} 
        \text{ and }
        \\
        \chi(C) 
        &
        := \begin{cases}
            \ell, & \text{if $C \in \{K_1, \Cstar_1 \setminus K_1 \}$,} \\
            \chistar(C), & \text{if $C \in \scriptC \cap \scriptCstar$.}
        \end{cases}
    \end{align*}
    Let $E_1$ be the set of edges that cross $K_1$ and $\Cstar_1 \setminus K_1$. Observe that each edge in $E_1$ is clustered together in $(\scriptCstar,\chistar)$ but separated in $(\scriptC,\chi)$; each edge not in $E_1$ has the same statuses in both solutions (i.e., for any edge not in $E_1$, if it is separated in one solution, it is also separated in the other; and if it is clustered by some color in one solution, it is also clustered by the same color in the other).
    We can therefore observe that,
    \begin{align*}
        &\text{for } e \in E_1,&& \cost_{(\scriptCstar,\chistar)}(e)=1-w^\ell(e) \;\text{ and }\; \cost_{(\scriptC,\chi)}(e)=1-w^-(e);\\
        &\text{for } e \notin E_1,&& \cost_{(\scriptCstar,\chistar)}(e)=\cost_{(\scriptC,\chi)}(e).
    \end{align*}
    Since $(\scriptCstar,\chistar)$ is an optimal solution, we have that 
    \begin{align}
        \sum_{e \in E_1}\cost_{(\scriptCstar,\chistar)}(e) \le \sum_{e \in E_1}\cost_{(\scriptC,\chi)}(e) \Longleftrightarrow \sum_{e \in E_1}w^-(e) \le \sum_{e \in E_1}w^\ell(e).\label{eq:case0-optimality_weighted}
    \end{align}
    For any $u \in K_1$, note that
    \begin{align*}
        w^{-}(u,\Cstar_1\setminus K_1)
        &
        = |\Cstar_1\setminus K_1| - w^{+}(u,\Cstar_1\setminus K_1)
        = |\Cstar_1\setminus K| - w^{+}(u,\Cstar_1\setminus K)
        \\&
        \ge |\Cstar_1\setminus K| - w^{+}(u,V\setminus K)
        > |\Cstar_1\setminus K| - \eta |K|,
    \end{align*}
    where the first equality comes from the probability constraint, and the last inequality follows from Lemma~\ref{lem:precluster-disagreement-bound_weighted}.
    We can thus derive that 
    \begin{equation}\label{eq:case0-LHS_weighted}
        \sum_{e \in E_1}w^-(e) > |K_1| \cdot (|\Cstar_1\setminus K| - \eta |K|)
        \geq |K_1| \cdot \eta |K|.
    \end{equation}
    
    On the other hand, for any $u \in K_1$, we have
    \begin{align*}
        w^{\ell}(u,\Cstar_1\setminus K_1) = w^{\ell}(u,\Cstar_1\setminus K) \le w^{+}(u,V\setminus K)< \eta |K|
    \end{align*}
    where the last inequality is again due to Lemma~\ref{lem:precluster-disagreement-bound_weighted}, yielding that
    \begin{equation*}
        \sum_{e \in E_1}w^\ell(e) < |K_1| \cdot \eta |K|.
    \end{equation*}
    This contradicts \eqref{eq:case0-optimality_weighted} due to \eqref{eq:case0-LHS_weighted}.
\end{proof}

\begin{lemma}[Restatement of Lemma~\ref{lem-main:precluster-subdivide} for \WCCC]\label{lem:precluster-subdivide_weighted}
    $\preclustering$ subdivides $\scriptCstar$, i.e., for any $K\in\preclustering$, there is $\Cstar \in \scriptCstar$ such that $K \subseteq \Cstar$.
\end{lemma}
\begin{proof}
    Suppose toward contradiction that $\scriptCstar$ is not subdivided by $\preclustering$.
    Then, there exists a non-singleton precluster $K \in \preclustering_{>1}$ such that $K \setminus \Cstar$ is nonempty for every cluster $\Cstar \in \scriptCstar$.
    We choose an arbitrary such precluster $K \in \preclustering_{>1}$, and let $\ell:=\colorpreclustering(K)$ be the color of $K$ in the preclustering.
    Let $\xi$ be a constant between $\frac{1+5\eta}{2}$ and $1-4\eta$.
    Note that we can always choose such $\xi$ because $\frac{1+5\eta}{2}< 1-4\eta$ for $\eta \in (0,\frac{1}{13})$.
    Observe that $\frac{1}{2}<\xi< 1$.

    We break the analysis into two cases depending on the existence of a cluster $\Cstar \in \scriptCstar$ that ``dominantly'' intersects $K$, i.e., $|K \cap \Cstar| \geq \xi |K|$.
    The overall structure of the proof for each case is similar to the proof of Lemma~\ref{lem:precluster-domination_weighted}: we respectively construct a solution whose cost is strictly less than $(\scriptCstar, \chistar)$, leading to contradiction that $(\scriptCstar, \chistar)$ is optimal.
    
    \textbf{Case 1. No dominant intersections.} Suppose that, for all $\Cstar \in \scriptCstar$, $|K \cap \Cstar| < \xi |K|$.
    Consider a new solution $(\scriptC',\chi')$ constructed by inserting $K$ colored $\ell = \colorpreclustering(K)$. More precisely, we have
    \begin{align*}
        \scriptC' &:= \{ K \} \cup \{\Cstar \setminus K \mid \Cstar \in \scriptCstar \text{ with } |\Cstar \setminus K| > 0\}
        \quad\text{ and }\quad
        \\
        \chi'(C) &:= \begin{cases}
            \ell, & \text{if $C = K$}; \\
            \chistar(\Cstar), & \text{if $C = \Cstar \setminus K$.}
        \end{cases}
    \end{align*}
    For $e=uv$ such that there exists $\Cstar \in \scriptCstar$ with $u,v\in \Cstar$, let $\Cstar_e$ (and $\Cstar_u$) denote the cluster in $\scriptCstar$ that contains both $u$ and $v$ (and $u$, respectively).
    We define $E_1$ and $E_2$ as follows.
    \begin{itemize}
        \item Let $E_1$ be the set of edges that are within some cluster in $\scriptCstar$ but across $K$ and $V \setminus K$.
        \item Let $E_2$ be the set of edges that are across clusters in $\scriptCstar$ but within $K$.
    \end{itemize}
    Note that $E_1$ and $E_2$ are disjoint.
    Observe that each edge in $E_1$ is clustered together in $(\scriptCstar,\chistar)$ but separated in $(\scriptC',\chi')$; each edge in $E_2$ is separated in $(\scriptCstar,\chistar)$ but clustered together in $(\scriptC',\chi')$; each edge not in $E_1\cup E_2$ has the same statuses in both solutions.
    Therefore, 
    \begin{align*}
        &\text{for } e \in E_1,&& \cost_{(\scriptCstar,\chistar)}(e)=1-w^{\chistar(\Cstar_e)}(e) \;\text{ and }\; \cost_{(\scriptC',\chi')}(e)=1-w^-(e),\\
        &\text{for } e \in E_2,&& \cost_{(\scriptCstar,\chistar)}(e)=1-w^-(e) \;\text{ and }\; \cost_{(\scriptC',\chi')}(e)=1-w^\ell(e),\\
        &\text{for } e \notin E_1\cup E_2,&& \cost_{(\scriptCstar,\chistar)}(e)=\cost_{(\scriptC',\chi')}(e).
    \end{align*}
    Since $(\scriptCstar,\chistar)$ is an optimal solution, we have that 
    \begin{align}
        &
        \sum_{e \in E_1}\cost_{(\scriptCstar,\chistar)}(e)+\sum_{e \in E_2}\cost_{(\scriptCstar,\chistar)}(e) 
        \le \sum_{e \in E_1}\cost_{(\scriptC',\chi')}(e)+\sum_{e \in E_2}\cost_{(\scriptC',\chi')}(e)
        \nonumber
        \\ 
        &
        \iff \sum_{e \in E_1}w^-(e)+\sum_{e \in E_2}w^\ell(e) 
        \le \sum_{e \in E_1}w^{\chistar(\Cstar_e)}(e)+\sum_{e \in E_2}w^-(e).\label{eq:case1-optimality_weighted}
    \end{align}

    Let us first lower-bound the left-hand side of \eqref{eq:case1-optimality_weighted}.
    For any $u \in K$, note that 
    \begin{align*}
        w^{\ell}(u,K\setminus \Cstar_u)
        &
        = |K\setminus \Cstar_u| - w^{-\ell}(u,K \setminus \Cstar_u)
        \ge |K\setminus \Cstar_u| - w^{-\ell}(u,K)
        \\&
        > |K\setminus \Cstar_u|- \eta |K|
        \ge
        (1-\xi-\eta)|K|,
    \end{align*}
    where the equality comes from the probability constraint, the second inequality from Lemma~\ref{lem:precluster-disagreement-bound_weighted}, and the last inequality from the condition of this case.
    Hence, 
    \begin{align}
        \sum_{e \in E_2}w^\ell(e) = \frac{1}{2}\sum_{u \in K}w^\ell(u,K\setminus \Cstar_u) > \frac{1 - \xi - \eta}{2} |K|^2.\label{eq:case1-LHS_weighted}
    \end{align}
    
    We now turn to bounding from above the right-hand side of \eqref{eq:case1-optimality_weighted}.
    For any $u \in K$, we have 
    \begin{align*}
        w^{\chistar(\Cstar_u)}(u,\Cstar_u\setminus K)
        &
        \le w^{+}(u,V\setminus K) < \eta |K|, \text{ and}
        \\
        w^-(u,K\setminus \Cstar_u) 
        &
        \le w^{-\ell}(u,K) < \eta |K|,        
    \end{align*}
    where the last inequalities follow from Lemma~\ref{lem:precluster-disagreement-bound_weighted}.
    Therefore,
    \begin{align}
        &
        \sum_{e \in E_1}w^{\chistar(\Cstar_e)}(e)+\sum_{e \in E_2}w^-(e) \nonumber
        \\&
        = \sum_{u \in K}w^{\chistar(\Cstar_u)}(u,\Cstar_u\setminus K)+ \frac{1}{2}\sum_{u \in K}w^{-}(u,K\setminus \Cstar_u) \nonumber
        \\&
        < 
        \frac{3}{2}\eta |K|^2.\label{eq:case1-RHS_weighted}
    \end{align}
    Since $\xi\le 1-4\eta$, we have
    \[
        \frac{1-\xi-\eta}{2} |K|^2 \geq \frac{3\eta}{2}|K|^2,
    \]
    leading to a contradiction to \eqref{eq:case1-optimality_weighted} because of \eqref{eq:case1-LHS_weighted} and \eqref{eq:case1-RHS_weighted}.

    \textbf{Case 2. A dominant intersection exists.}
    Suppose now there exists a cluster $\Cstar_1 \in \scriptCstar$ such that $|K \cap \Cstar_1| \ge \xi |K|$.
    Note that we always have at most one such cluster since $\xi>\frac{1}{2}$.
    Let $K_1:=K \cap \Cstar_1$.
    
    We first claim that $\chistar(\Cstar_1) \neq \colorpreclustering(K)$.
    Otherwise, if $\chistar(\Cstar_1) = \colorpreclustering(K)$, we construct a solution $(\scriptC'',\chi'')$ by transferring one arbitrary vertex $u \in K \setminus \Cstar_1$ to $\Cstar_1$.
    In other words,
    \begin{align*}
        \scriptC'' &:= \{ \Cstar_1 \cup \{u\} \} \cup \{ \Cstar \setminus \{u\} \mid \Cstar \in \scriptCstar \setminus \{\Cstar_1\} \}
        \text{ and}
        \\
        \chi''(C) &:= \begin{cases}
            \chistar(\Cstar_1), & \text{if $C = \Cstar_1 \cup \{u\}$,} \\
            \chistar(\Cstar), & \text{if $C = \Cstar \setminus \{u\}$.}
        \end{cases}
    \end{align*}
    Since $K \setminus \Cstar_1$ is nonempty, we can always choose $u \in K \setminus \Cstar_1$ in the above construction.
    Recall also that $\ell = \colorpreclustering(K) = \chistar(\Cstar_1)$. 
    
    We again use $\Cstar_u$ to denote the cluster in $\scriptCstar$ that contains $u$. Let us define $E_1$ and $E_2$ as follows.
    \begin{itemize}
        \item Let $E_1$ be the set of edges that are incident to $u$ and within $\Cstar_u$.
        \item Let $E_2$ be the set of edges that are incident to $u$ and within $\Cstar_1 \cup \{u\}$.
    \end{itemize}
    Note that $E_1$ and $E_2$ are disjoint.
    Observe that each edge in $E_1$ is clustered together in $(\scriptCstar,\chistar)$ but separated in $(\scriptC'',\chi'')$ while each edge in $E_2$ is separated in $(\scriptCstar,\chistar)$ but clustered together in $(\scriptC'',\chi'')$.
    Each edge not in $E_1\cup E_2$ has the same statuses in both solutions.
    Therefore, 
    \begin{align*}
        &\text{for } e \in E_1,&& \cost_{(\scriptCstar,\chistar)}(e)=1-w^{\chistar(\Cstar_u)}(e) \text{ and } \cost_{(\scriptC'',\chi'')}(e)=1-w^-(e),\\
        &\text{for } e \in E_2,&& \cost_{(\scriptCstar,\chistar)}(e)=1-w^-(e) \text{ and } \cost_{(\scriptC'',\chi'')}(e)=1-w^\ell(e),\\
        &\text{for } e \notin E_1\cup E_2,&& \cost_{(\scriptCstar,\chistar)}(e)=\cost_{(\scriptC'',\chi'')}(e).
    \end{align*}
    Since $(\scriptCstar,\chistar)$ is an optimal solution, we have that 
    \begin{align}
        &\sum_{e \in E_1}\cost_{(\scriptCstar,\chistar)}(e)+\sum_{e \in E_2}\cost_{(\scriptCstar,\chistar)}(e) 
        \le \sum_{e \in E_1}\cost_{(\scriptC'',\chi'')}(e)+\sum_{e \in E_2}\cost_{(\scriptC'',\chi'')}(e) \nonumber
        \\ 
        &
        \iff\sum_{e \in E_1}w^-(e)+\sum_{e \in E_2}w^\ell(e) 
        \le \sum_{e \in E_1}w^{\chistar(\Cstar_u)}(e)+\sum_{e \in E_2}w^-(e).\label{eq:case2-1-optimality_weighted}
    \end{align}
    
    For a lower bound on the left-hand side of \eqref{eq:case2-1-optimality_weighted}, note that 
    \begin{align}
        \sum_{e \in E_2}w^\ell(e) 
        &
        = w^{\ell}(u,\Cstar_1)
        \ge w^{\ell}(u,K_1)
        =|K_1|-w^{-\ell}(u,K_1) \nonumber
        \\&
        \ge |K_1|-w^{-\ell}(u,K)
        > |K_1|-\eta |K| 
        \ge 
        (\xi - \eta) |K|,\label{eq:case2-1-LHS_weighted}
    \end{align}
    where the second equality is due to the probability constraint, the second-to-last inequality is from Lemma~\ref{lem:precluster-disagreement-bound_weighted}, and the last inequality is due to the condition of the case.

    On the other hand, for an upper bound on the right-hand side of \eqref{eq:case2-1-optimality_weighted}, we have
    \begin{align}
        \sum_{e \in E_1}w^{\chistar(\Cstar_u)}(e)
        &
        = w^{\chistar(\Cstar_u)}(u, \Cstar_u)
        \le w^{\chistar(\Cstar_u)}(u, V \setminus \Cstar_1)
        \le w^{+}(u, V \setminus \Cstar_1)
        \notag
        \\&
        = w^{+}(u, K \setminus \Cstar_1) + w^{+}(u,(V\setminus K)\setminus \Cstar_1) \notag
        \\&
        \leq |K \setminus \Cstar_1| + w^{+}(u,V\setminus K) \nonumber
        \\&
        \le (1-\xi)|K|+w^{+}(u,V\setminus K) \notag
        \\&
        < 
        (1 - \xi + \eta) |K|, \label{eq:case2-1-RHS-2_weighted}
    \end{align}
    where the second equality follows from the probability constraint, the second-to-last inequality from the condition that $|K \cap \Cstar_1| \geq \xi |K|$, and the last inequality from Lemma~\ref{lem:precluster-disagreement-bound_weighted}.
    We also have
    \begin{align}
        \sum_{e \in E_2}w^-(e)
        & 
        = w^{-}(u,\Cstar_1)
        \le w^{-\ell}(u,\Cstar_1)
        \leq w^{-\ell}(u, K) + w^{-\ell}(u, \Cstar_1 \setminus K) \notag
        \\&
        \le w^{-\ell}(u,K)+|\Cstar_1\setminus K|
        < \eta |K| + 2\eta |K| = 3\eta |K|, \label{eq:case2-1-RHS-1_weighted}
    \end{align}
    where the last inequality comes from Lemma~\ref{lem:precluster-disagreement-bound_weighted} and Lemma~\ref{lem:precluster-domination_weighted}.
    Observe that
    \[
        (\xi - \eta) |K| \geq (1 - \xi + 4\eta) |K|
    \]
    due to the choice of $\xi \geq \frac{1 + 5\eta}{2}$, yielding a contradiction to \eqref{eq:case2-1-optimality_weighted} due to \eqref{eq:case2-1-LHS_weighted}, \eqref{eq:case2-1-RHS-2_weighted}, and \eqref{eq:case2-1-RHS-1_weighted}.
    This completes the proof of our claim that $\chistar(\Cstar_1) \neq \colorpreclustering(K)$.

    Let $\ell':=\chistar(\Cstar_1) \neq \ell$.
    We now construct another solution $(\scriptC, \chi)$ from $(\scriptCstar, \chistar)$ by splitting $\Cstar_1$ into $K_1$ colored $\ell = \colorpreclustering(K)$ and $\Cstar_1 \setminus K_1$ colored $\ell' = \chistar(\Cstar_1)$, i.e.,
    \[
        \scriptC := \scriptCstar \setminus \{\Cstar_1\} \cup \{ K_1, \Cstar_1 \setminus K_1 \} 
        \text{ and }
        \chi(C) := \begin{cases}
            \ell, & \text{if $C = K_1$,} \\
            \chistar(\Cstar_1), & \text{if $C = \Cstar_1 \setminus K_1$,} \\
            \chistar(C), & \text{if $C \in \scriptC \cap \scriptCstar$.}
        \end{cases}.
    \]
    Let us define $E_1$ and $E_3$ as follows.
    \begin{itemize}
        \item Let $E_1$ be the set of edges that cross $K_1$ and $\Cstar_1\setminus K_1$.
        \item Let $E_3$ be the set of edges that are within $K_1$.
    \end{itemize}
    Note that $E_1$ and $E_3$ are disjoint.
    Observe that each edge in $E_1$ is clustered together in $(\scriptCstar,\chistar)$ but separated in $(\scriptC,\chi)$ while each edge in $E_3$ is clustered together by color $\ell'$ in $(\scriptCstar,\chistar)$ but clustered together by color $\ell$ in $(\scriptC,\chi)$. Each edge not in $E_1\cup E_3$ has the same statuses in both solutions.
    Therefore, 
    \begin{align*}
        &\text{for } e \in E_1,&& \cost_{(\scriptCstar,\chistar)}(e)=1-w^{\ell'}(e) \;\text{ and }\; \cost_{(\scriptC,\chi)}(e)=1-w^-(e),\\
        &\text{for } e \in E_3,&& \cost_{(\scriptCstar,\chistar)}(e)=1-w^{\ell'}(e) \;\text{ and }\; \cost_{(\scriptC,\chi)}(e)=1-w^\ell(e),\\
        &\text{for } e \notin E_1\cup E_3,&& \cost_{(\scriptCstar,\chistar)}(e)=\cost_{(\scriptC,\chi)}(e).
    \end{align*}
    Since $(\scriptCstar,\chistar)$ is an optimal solution, we have that 
    \begin{align}
        &\sum_{e \in E_1}\cost_{(\scriptCstar,\chistar)}(e)+\sum_{e \in E_3}\cost_{(\scriptCstar,\chistar)}(e) 
        \le \sum_{e \in E_1}\cost_{(\scriptC,\chi)}(e)+\sum_{e \in E_3}\cost_{(\scriptC,\chi)}(e) \notag
        \\ 
        &\iff \sum_{e \in E_1}w^-(e)+\sum_{e \in E_3}w^\ell(e) 
        \le \sum_{e \in E_1}w^{\ell'}(e)+\sum_{e \in E_3}w^{\ell'}(e).\label{eq:case2-2-optimality_weighted}
    \end{align}
    
    To find a lower bound on the left-hand side of \eqref{eq:case2-2-optimality_weighted}, consider any $u \in K_1$:
    \begin{align*}
        w^\ell(u,K_1)
        &
        = |K_1|-1 - w^{-\ell}(u,K_1)
        \ge |K_1|-1 - w^{-\ell}(u,K)
        \\&
        > |K_1|-1 - \eta (|K|-1) 
        \ge \xi |K| -1  - \eta (|K|-1),
    \end{align*}
    where the second inequality comes from Lemma~\ref{lem:precluster-disagreement-bound_weighted}, and the last inequality comes from the condition of $K_1$.
    Hence,
    \begin{align}
        \sum_{e \in E_3}w^\ell(e) = \frac{1}{2} \sum_{u \in K_1} w^\ell(u, K_1)
        > \frac{1}{2}|K_1|\cdot(\xi|K|-1 -\eta(|K|-1)).\label{eq:case2-2-LHS_weighted}
    \end{align}

    We now upper-bound the right-hand side of \eqref{eq:case2-2-optimality_weighted}.
    For any $u \in K_1$, we have
    \begin{align*}
        w^{\ell'}(u,\Cstar_1 \setminus K_1)
        &
        \le w^+(u,\Cstar_1 \setminus K_1)
        = w^+(u,\Cstar_1 \setminus K)
        \\&
        \le w^+(u,V \setminus K)
        < \eta (|K|-1),
    \end{align*}
    where the last inequality is due to Lemma~\ref{lem:precluster-disagreement-bound_weighted}. This shows that
    \begin{align}
        \sum_{e \in E_1}w^{\ell'}(e) = \sum_{u \in K_1}w^{\ell'}(u,\Cstar_1 \setminus K_1) < |K_1|\cdot \eta (|K|-1). \label{eq:case2-2-RHS-1_weighted}
    \end{align}
    Lastly, for every $u \in K_1$,
    \begin{align*}
        w^{\ell'}(u,K_1) \le w^{-\ell}(u,K_1) \le w^{-\ell}(u,K) <\eta (|K|-1)
    \end{align*}
    where the last inequality comes from Lemma~\ref{lem:precluster-disagreement-bound_weighted}, yielding
    \begin{align}
        \sum_{e \in E_3}w^{\ell'}(e) = \frac{1}{2} \sum_{u \in K_1} w^{\ell'}(u,K_1)
        < \frac{1}{2}|K_1|\cdot\eta (|K|-1).\label{eq:case2-2-RHS-2_weighted}
    \end{align}

    We claim that
    \[
        \frac{1}{2}|K_1|\cdot(\xi|K|-1 -\eta(|K|-1))
        >
        \frac{3}{2}|K_1|\cdot\eta (|K|-1).
    \]
    If the claim is true, we can complete the proof because it yields a contradiction to \eqref{eq:case2-2-optimality_weighted} due to \eqref{eq:case2-2-LHS_weighted}, \eqref{eq:case2-2-RHS-1_weighted}, and \eqref{eq:case2-2-RHS-2_weighted}.
    It suffices to show that $\xi|K|-1 - 4\eta(|K|-1) > 0$.
    Observe that $\xi > \frac{1+5\eta}{2} > \frac{1}{2} > \frac{4}{13} > 4\eta$ by the choice of $\xi$ and $\eta$.
    We can thus see
    \[
        \xi|K|-1 -4\eta(|K|-1)= (\xi-4\eta)|K|-1+4\eta \ge 2\xi-4\eta-1 > \eta,
    \]
    where the first inequality comes from the fact that $K$ is non-singleton and $\xi > 4\eta$, and the last inequality from the fact that $\xi > \frac{1 + 5 \eta}{2}$.
\end{proof}

The above lemma shows that each precluster $K \in \preclustering$ is fully contained in some cluster $\Cstar \in \scriptCstar$. 
We can also show that every non-singleton precluster must be colored the same as the cluster in $\scriptCstar$ containing it.

\begin{lemma}[Restatement of Lemma~\ref{lem-main:precluster-unique} for \WCCC]\label{cor:precluster-unique_weighted}
    For every cluster $\Cstar \in \scriptCstar$ and any non-singleton precluster $K \in \preclustering$ contained in $\Cstar$, we have $\colorpreclustering(K) =  \chistar(\Cstar)$.
\end{lemma}
\begin{proof}
    Suppose toward contradiction that there exists a non-singleton precluster $K$ and a cluster $\Cstar_1$ such that $K \subseteq \Cstar_1$ but $\colorpreclustering(K) \neq \chistar(\Cstar_1)$. We can show that a solution $(\scriptC, \chi)$ obtained by splitting $\Cstar_1$ into $K$ of color $\colorpreclustering(K)$ and $\Cstar_1 \setminus K$ of color $\chistar(\Cstar_1)$ has smaller cost than that of $(\scriptCstar, \chistar)$.
    The proof indeed follows from exactly the same argument as in the latter half of Case 2 in the proof of Lemma~\ref{lem:precluster-subdivide_weighted}.
\end{proof}

\begin{lemma}[Restatement of Lemma~\ref{lem:precluster-constant-approx} for \WCCC]\label{lem:precluster-constant-approx_weighted}
    $(\preclustering, \colorpreclustering)$ is a constant-approximate solution.
\end{lemma}
\begin{proof}
    We show that the cost of $(\preclustering, \colorpreclustering)$ is within a constant factor of that of $(\initialclustering,\colorinitialclustering)$; since $(\initialclustering,\colorinitialclustering)$ is a constant-approximate solution, this completes the proof.
    Note moreover that, for every edge $e \in E$, if the cost incurred by $e$ in $(\preclustering, \colorpreclustering)$ changes from $(\initialclustering, \colorinitialclustering)$, this must be because at least one endpoint of $e$ is contained in a non-singleton cluster in $\initialclustering$, but becomes singleton in $\preclustering$.
    Hence, for some absolute constant $c$, if we can show that, for every non-singleton cluster $C \in \initialclustering$, 
    \begin{align}
        & \sum_{\substack{u \in C:\\\{u\}\in \preclustering}}\sum_{v \in V\setminus \{u\}}\left(\cost_{(\preclustering, \colorpreclustering)}(uv) - \cost_{(\initialclustering,\colorinitialclustering)}(uv)\right) 
        \nonumber \\
        & \quad \le c \sum_{\substack{u \in C:\\\{u\}\in \preclustering}}\sum_{v \in V\setminus \{u\}}\cost_{(\initialclustering,\colorinitialclustering)}(uv), \label{eq:preclu:approx:weighted}
    \end{align}
    this immediately completes the proof of this lemma since every edge is incident to vertices from at most two clusters in $\initialclustering$.
    
    Consider any non-singleton cluster $C \in \initialclustering$ of color $\ell:=\colorinitialclustering(C)$.
    Observe that, if a vertex $u\in C$ is a singleton precluster in $\preclustering$, we have that
    \begin{align*}
        \sum_{v \in V\setminus \{u\}}\cost_{(\preclustering, \colorpreclustering)}(uv)
        & = w^{+}(u,C) + w^{+}(u,V\setminus C), \text{ and}
        \\
        \sum_{v \in V\setminus \{u\}}\cost_{(\initialclustering,\colorinitialclustering)}(uv)
        & = w^{-\ell}(u,C) + w^{+}(u,V\setminus C).
    \end{align*}
    Hence, \eqref{eq:preclu:approx:weighted} is equivalent to
    \begin{equation} \label{eq:preclu:approx:weighted02}
        \sum_{\substack{u \in C:\\\{u\}\in \preclustering}}\left(w^{+}(u,C)-w^{-\ell}(u,C)\right)
        \le c \sum_{\substack{u \in C:\\\{u\}\in \preclustering}}\left(w^{-\ell}(u,C) + w^{+}(u,V\setminus C)\right).
    \end{equation}
    In fact, we show a stronger inequality:
    \[
        \sum_{\substack{u \in C:\\\{u\}\in \preclustering}}w^{\ell}(u,C)
        \le c \sum_{\substack{u \in C:\\\{u\}\in \preclustering}}\left(w^{-\ell}(u,C) + w^{+}(u,V\setminus C)\right).
    \]
    
    Let
    \begin{align*}
        B_1 &:= \{ u \in C \mid w^{-\ell}(u,C) \ge \alpha (|C|-1)\} 
        \;\text{ and }\;
        \\
        B_2 &:= \{ u \in C \mid w^{+}(u,V\setminus C) \ge \alpha (|C|-1)\}.
    \end{align*}
    Note that $B_1 \cup B_2$ is the set of marked vertices in $C$ right after the first iteration over $V$.
    We break the analysis into two cases depending on the size of $B_1 \cup B_2$.
    
    \textbf{Case 1. $|B_1\cup B_2| < \beta(|C|-1)$.} 
    In this case, the vertices in $C \setminus (B_1 \cup B_2)$ remain unmarked. We show that for any $u \in B_1 \cup B_2$, $w^{\ell}(u,C)\le \frac{1}{\alpha} (w^{-\ell}(u,C) + w^{+}(u,V\setminus C) )$.
    For any $u \in B_1$, observe that
    \begin{align*}
        w^\ell(u,C)
        &
        =|C|-1-w^{-\ell}(u,C)
        \le (1 - \alpha) (|C|-1)
        \\&
        \le \frac{1-\alpha}{\alpha}w^{-\ell}(u,C)
        <\frac{1}{\alpha}w^{-\ell}(u,C),
    \end{align*}
    where the first and second inequalities are due to the definition of $B_1$; on the other hand, for $u \in B_2$, we have
    \begin{align*}
        w^\ell(u,C) \le |C|-1 \le \frac{1}{\alpha}w^{+}(u,V\setminus C),
    \end{align*}
    where the last inequality is due to the definition of $B_2$.
    
    \textbf{Case 2. $|B_1\cup B_2| \ge \beta(|C|-1)$.} 
    Recall that, in this case, the algorithm marks all vertices in $C$. Therefore,
    \[
        \sum_{\substack{u \in C:\\\{u\}\in \preclustering}}w^{\ell}(u,C) 
        \le |C|(|C|-1)
        \leq 2(|C|-1)^2,
    \]
    where the first inequality is due to the probability constraints, and the last inequality is because $C$ is non-singleton. 
    Note that, by the definition of $B_1$,
    \[
        \sum_{u \in C}w^{-\ell}(u,C) 
        \ge \sum_{u \in B_1}w^{-\ell}(u,C)
        \ge |B_1| \cdot \alpha (|C|-1),
    \]
    and, by the definition of $B_2$, 
    \[
        \sum_{u \in C}w^{+}(u,V\setminus C)
        \ge \sum_{u \in B_2}w^{+}(u,V\setminus C)
        \ge |B_2|\cdot \alpha (|C|-1).
    \]
    We can thus derive that
    \[
        \sum_{u \in C}\left(w^{-\ell}(u,C) + w^{+}(u,V\setminus C)\right)
        \ge \alpha (|B_1| + |B_2|) (|C| - 1)
        \ge \alpha \beta (|C|-1)^2,
    \]
    where the last inequality is from the condition of this case.

    Hence, choosing $c = \frac{2}{\alpha \beta}$ is sufficient for \eqref{eq:preclu:approx:weighted02}, completing the proof.
\end{proof}

\subsection{Constructing \texorpdfstring{$\Eadm$}{E adm}}\label{app:preclu:adm}
In the previous subsections, we have presented the construction of our preclustering $\preclustering$ and its color assignment $\colorpreclustering$ for \CCC and \WCCC, respectively.
This subsection is devoted to constructing a set $\Eadm$ of admissible edges satisfying Theorem~\ref{thm:preclu:main} with $\preclustering$ and $\colorpreclustering$.
To this end, we first identify a nearly optimal solution respecting the preclustered instance with a small-sized $\Eadm$.

For \CCC, for any $S,S'\subseteq V$, let $w^+(S,S'):=|\{uv \in E^+ \mid u\in S,v\in S' \}|$ and let $w^-(S,S'):=|\{uv \in E^- \mid u\in S,v\in S' \}|$; for \WCCC, the argument below follows in the same way with the notation defined for \WCCC in Section~\ref{sec_def}.
The rest of the notation is identical across the two problems.
Let $\preclustering_C :=\{K \in \preclustering\mid K \subseteq C\}$ for a cluster $C \subseteq V$.
Fix an optimal solution $(\scriptCstar,\chistar)$. 
Given $\varepsilon >0$, let us construct a nearly optimal solution $(\scriptCstarnear,\chistarnear)$ as follows.
Initially, we set $(\scriptCstarnear,\chistarnear):=(\scriptCstar,\chistar)$. If there exists a cluster $C\in\scriptCstarnear$ and a precluster $K \in \preclustering_C$ such that $|\preclustering_C|>1$ and 
\[
    w^+(K,C\setminus K) - w^-(K, C\setminus K) \le 2\varepsilon\cdot (w^+(K,V\setminus K)+w^-(K,K)),
\]
we update $\scriptCstarnear$ by splitting $C$ into $K$ and $C\setminus K$ with both $\chistarnear(K)$ and $\chistarnear(C\setminus K)$ being $\chistarnear(C)$.
We repeat this process until no such $C\in\scriptCstarnear$ and $K \in \preclustering$ exist.

Since $(\scriptCstar, \chistar)$ is initially subdivided by $\preclustering$ due to Lemma~\ref{lem:precluster-subdivide} (or Lemma~\ref{lem:precluster-subdivide_weighted} for \WCCC) and the algorithm also respects $\preclustering$ in every iteration, it is clear that $\scriptCstarnear$ is subdivided by $\preclustering$.
We now show that this solution is nearly optimal.

\begin{lemma}
    $(\scriptCstarnear,\chistarnear)$ is a $(1+O(\varepsilon))$-approximate solution.
\end{lemma}
\begin{proof}
    We show that, for each cluster $C \in \scriptCstar$ in the initial optimal solution, the ``net increase'' in cost of the modified solution $\scriptCstarnear$ due to the preclusters contained in $C$ during the algorithm is bounded by a factor $O(\varepsilon)$ of the cost incurred by $C$ in the initial $\scriptCstar$.
    
    Let $K_1, K_2, \ldots, K_t \subseteq C$ be the preclusters split off during the algorithm in the same order (e.g., $K_1$ is split off first, $K_2$ is then split off next, and so forth).
    For clarity, let $C_0 := C$, and for $i \in \{1, \ldots, t\}$, we denote
    \begin{itemize}
        \item by $C_i := C_{i-1} \setminus K_i$ the remaining cluster after $K_i$ is split off; and
        \item by $P_i := \bigcup_{j=1}^i K_j$ the preclusters that have been split off until $K_i$ is split off (included).
    \end{itemize}
    Let $P_0 := \emptyset$ for consistency.
    
    Observe that, for every $i \in \{1, \ldots, t\}$, when the algorithm splits $C_{i-1}$ into $K_i$ and $C_i$, the $+$ edges across $K_i$ and $C_i$ may become disagreed, while the $-$ edges across $K_i$ and $C_i$ become agreed.
    Therefore, the net increase in cost is at most $w^+(K_i, C_i) - w^-(K_i, C_i)$, implying that the total net increase in cost of $\scriptCstarnear$ is bounded from above by
    \begin{equation} \label{expr:adm:netinc}
        \UBstarnear := \sum_{i = 1}^t \left( w^+(K_i, C_i) - w^-(K_i, C_i) \right).
    \end{equation}
    Note also that, by the execution in each iteration where $K_i$ is split off, we can bound each net increase by
    \begin{align}
        &
        w^+(K_i, C_i) - w^-(K_i, C_i)
        \nonumber \\&
        = w^+(K_i, C_{i-1} \setminus K_i) - w^-(K_i, C_{i-1} \setminus K_i)
        \nonumber \\&
        \leq 2 \varepsilon \left( w^+(K_i, V \setminus K_i) + w^-(K_i, K_i) \right). \label{ineq:adm:splitcond}
    \end{align}

    We want to bound this net increase in cost of $\scriptCstarnear$ by the cost incurred due to $C_0$ in $\scriptCstar$.
    For $i \in \{1, \ldots, t\}$, each $-$ edge $uv \in E^-$ with $u \in K_i$ and $v \in C_0$ is a disagreement in the initial solution $\scriptCstar$.
    Similarly, each $+$ edge $uv \in E^+$ with $u \in K_i$ and $v \in V \setminus C_0$ is also disagreed in $\scriptCstar$ regardless of its color.
    We thus aim at bounding $\UBstarnear$ in \eqref{expr:adm:netinc} by
    \begin{equation} \label{expr:adm:initcost}
        \LBstar := \sum_{i = 1}^t \left( w^+(K_i, V \setminus C_0) + w^-(K_i, C_0) \right).
    \end{equation}
    Observe that, for each $i \in \{1, \ldots, t\}$,
    \begin{align*}
        &
        w^+(K_i, V \setminus C_0) + w^-(K_i, C_0)
        \\&
        = \left( w^+(K_i, V \setminus K_i) - w^+(K_i, C_i) - w^+(K_i, P_{i-1}) \right)
        \\&
        \qquad + \left( w^-(K_i, K_i) + w^-(K_i, C_i) + w^-(K_i, P_{i-1}) \right)
        \\&
        =  \left( w^+(K_i, V \setminus K_i) + w^-(K_i, K_i) \right)
        - \left( w^+(K_i, C_i) - w^-(K_i, C_i) \right)
        \\&
        \qquad - \left( w^+(K_i, P_{i-1}) - w^-(K_i, P_{i-1}) \right)
        \\&
        \geq (1 - 2\varepsilon) \left( w^+(K_i, V \setminus K_i) + w^-(K_i, K_i) \right) 
        \\&
        \qquad - \left( w^+(K_i, P_{i-1}) - w^-(K_i, P_{i-1}) \right),
        \nonumber
    \end{align*}
    where the inequality is due to Inequality~\eqref{ineq:adm:splitcond}.
    We can thus derive that
    \begin{align}
        &
        w^+(K_i, V \setminus K_i) + w^-(K_i, K_i)
        \nonumber \\&
        \leq \frac{1}{1-2\varepsilon} \left[ \left( w^+(K_i, V \setminus C_0) + w^-(K_i, C_0) \right) + \left( w^+(K_i, P_{i-1}) - w^-(K_i, P_{i-1}) \right) \right].
        \nonumber
    \end{align}
    By plugging this inequality back into Inequality~\eqref{ineq:adm:splitcond}, we have
    \begin{align}
        &
        w^+(K_i, C_i) - w^-(K_i, C_i)
        \nonumber \\&
        \leq \frac{2\varepsilon}{1-2\varepsilon} \left[ \left( w^+(K_i, V \setminus C_0) + w^-(K_i, C_0) \right) + \left( w^+(K_i, P_{i-1}) - w^-(K_i, P_{i-1}) \right) \right].
        \nonumber
    \end{align}
    Summing the above inequality over all $i \in \{1, \ldots, t\}$ yields
    \begin{equation} \label{ineq:eadm:ublb}
        \UBstarnear \leq \frac{2\varepsilon}{1 - 2 \varepsilon} \left[ \LBstar + \sum_{i=1}^t \left( w^+(K_i, P_{i-1}) - w^-(K_i, P_{i-1}) \right) \right]
    \end{equation}
    by the definitions of $\UBstarnear$ and $\LBstar$ in \eqref{expr:adm:netinc} and \eqref{expr:adm:initcost}, respectively.

    Let us further define
    \begin{align*}
        \KPstarnear
        &
        := \sum_{i=1}^t \left( w^+(K_i, P_{i-1}) - w^-(K_i, P_{i-1}) \right);
        \\ 
        \KCstarnear
        &
        := \sum_{i = 1}^t \left(  w^+(K_i, C_t) - w^-(K_i, C_t) \right).
    \end{align*}
    Observe that
    \begin{align}
        \UBstarnear
        & 
        = \sum_{i = 1}^t \left( w^+(K_i, C_i) - w^-(K_i, C_i) \right)
        \nonumber \\&
        = \sum_{i = 1}^t \left[ \left( w^+(K_i, C_i \setminus C_t) - w^-(K_i, C_i \setminus C_t) \right) + \left( w^+(K_i, C_t) - w^-(K_i, C_t) \right) \right]
        \nonumber \\&
        = \sum_{i = 1}^t \sum_{j = i+1}^t \left( w^+(K_i, K_j) - w^-(K_i, K_j) \right)
        + \sum_{i = 1}^t \left( w^+(K_i, C_t) - w^-(K_i, C_t) \right) 
        \nonumber \\&
        = \sum_{j = 1}^t \left( w^+(P_{j-1}, K_j) - w^-(P_{j-1}, K_j) \right)
        + \sum_{i = 1}^t \left( w^+(K_i, C_t) - w^-(K_i, C_t) \right) 
        \nonumber \\&
        = \KPstarnear + \KCstarnear.
        \nonumber
    \end{align}
    Note also that, due to the definition of $\LBstar$ in \eqref{expr:adm:initcost},
    \begin{equation*}
        - \KCstarnear 
        \leq \sum_{i=1}^t w^-(K_i, C_t) 
        \leq \sum_{i=1}^t w^-(K_i, C_0) 
        \leq \LBstar,
    \end{equation*}
    implying that
    \[
        \KPstarnear = \UBstarnear - \KCstarnear \leq \UBstarnear + \LBstar.
    \]
    We can therefore derive from Inequality~\eqref{ineq:eadm:ublb}
    \[
        \UBstarnear 
        \leq \frac{2\varepsilon}{1 - 2\varepsilon} \left( \LBstar + \KPstarnear \right) 
        \leq \frac{2\varepsilon}{1 - 2\varepsilon} \left( 2\LBstar + \UBstarnear \right).
    \]
    By rearranging the terms, we can obtain
    \[
        \UBstarnear \leq \frac{4\varepsilon}{1 - 4\varepsilon} \LBstar.
    \]
    Recall the definition of $\LBstar$ in \eqref{expr:adm:initcost}.
    For each $i \in \{1, \ldots, t\}$, we charge $\frac{4\varepsilon}{1 - 4\varepsilon}$ to each edge in $\{uv \in E^+ \mid u \in K_i \wedge v \in V \setminus C_0\} \cup \{uv \in E^- \mid u \in K_i \wedge v \in C_0\}$.
    This charge is sufficient to cover $\UBstarnear$, and hence, the net increase in cost due to each cluster $C_0 \in \scriptCstar$.
    When we process this charging over all clusters in $\scriptCstar$, we can observe that each edge disagreed in the initial $\scriptCstar$ is charged $\frac{4 \varepsilon}{1 - 4\varepsilon}$ at most twice from its endpoints, respectively.
    Hence, the total charged amount is within a factor $\frac{8\varepsilon}{1 - 4\varepsilon} = O(\varepsilon)$ of the cost of the initial solution $\scriptCstar$.
\end{proof}

Note that, for any cluster $C\in\scriptCstarnear$, each precluster $K \in \preclustering_C$ contained in $C$ satisfies
\begin{align*}
    &
    w^+(K,C\setminus K) - w^-(K, C\setminus K) 
    \\&
    =w^+(K,C\setminus K) - (|K||C\setminus K|-w^+(K,C\setminus K))
    \\&
    =2w^+(K,C\setminus K)-|K||C \setminus K|.
\end{align*}
For a cluster $C\in\scriptCstarnear$ with $|\preclustering_C|>1$, each precluster $K \in \preclustering_C$ satisfies 
\begin{align}
    2w^+(K,C\setminus K)-|K||C \setminus K| >2\varepsilon \cdot (w^+(K,V\setminus K)+w^-(K,K)) \ge 0.\label{eq:near-optimum-atom}
\end{align}
For notational simplicity, let us define
\[
    d(K):=\frac{w^+(K,V\setminus K)}{|K|}+\frac{|K|}{2}.
\]
Observe that
\begin{align}
    d(K)
    &
    = \frac{w^+(K,V\setminus K)}{|K|}+\frac{|K|}{2}
    \ge \frac{w^+(K,C\setminus K)}{|K|}+\frac{|K|}{2}
    \nonumber \\&
    > \frac{1}{2}\cdot \frac{|K||C \setminus K|}{|K|} +\frac{|K|}{2}
    = \frac{1}{2}(|C\setminus K|+|K|)= \frac{|C|}{2},\label{eq:atom-degree-lb}
\end{align}
where the last inequality is from Inequality~\eqref{eq:near-optimum-atom}. 
Similarly, we have the following:
\begin{align}
    \varepsilon \cdot d(K) &
    =\varepsilon \cdot \left( \frac{w^+(K,V\setminus K)}{|K|}+\frac{|K|}{2}\right)
    \nonumber\\&
    =\varepsilon \cdot \left(\frac{w^+(K,V\setminus K)+w^-(K,K)-w^-(K,K)}{|K|}+\frac{|K|}{2}\right)
    \nonumber\\&
    <\frac{1}{2|K|}\cdot\left(2w^+(K,C\setminus K)-|K||C \setminus K|\right)+\varepsilon \cdot \left(\frac{-w^-(K,K)}{|K|}+\frac{|K|}{2}\right)
    \nonumber\\&
    <\frac{w^+(K,C\setminus K)}{|K|}-\frac{|C \setminus K|}{2} + \frac{|K|}{2}\label{eq:atom-degree-ub-2}
    \\&
    \le\frac{|C \setminus K|}{2}+\frac{|K|}{2}=\frac{|C|}{2},\label{eq:atom-degree-ub-1}
\end{align} 
where the first inequality is again due to Inequality~\eqref{eq:near-optimum-atom}, and the last inequality follows from the fact that $w^+(K, C \setminus K) \leq |K||C\setminus K|$.
\begin{lemma}\label{lem:degree-similar}
    For a cluster $C\in\scriptCstarnear$ and two preclusters $K, K' \in \preclustering_C$, we have 
    \[
        \varepsilon \cdot d(K') < d(K) < \frac{1}{\varepsilon} \cdot d(K').
    \]
\end{lemma}
\begin{proof}
    It suffices to show $d(K) > \varepsilon\cdot d(K')$. It is trivial when $K=K'$. 
    On the other hand, if $K\neq K'$, Inequalities~\eqref{eq:atom-degree-lb} and \eqref{eq:atom-degree-ub-1} complete the proof since $|\preclustering_C|>1$.
\end{proof}

For a precluster $K \in \preclustering$, let $N_1(K):=\{K' \in \preclustering \mid \varepsilon \cdot d(K') < d(K) < \frac{1}{\varepsilon} \cdot d(K')\}$. Note that $K \in N_1(K)$. 
For distinct $K,K' \in \preclustering$ and $\preclustering'\subseteq \preclustering$ such that $K,K' \in \preclustering'$, we define
\begin{align*}
    W(K,K',\preclustering')
    & :=\sum_{K'' \in \preclustering' \setminus \{K,K'\}} \left[ |K''|\cdot\frac{w^+(K,K'')}{|K|\cdot |K''|}\cdot \frac{w^+(K',K'')}{|K'|\cdot |K''|} \right]
    \\&
    \qquad +\frac{w^+(K,K')}{|K|}+\frac{w^+(K',K)}{|K'|}
\end{align*}
\begin{lemma} \label{lem:preclu:W}
    For a cluster $C\in\scriptCstarnear$ with $|\preclustering_C|>1$ and two distinct preclusters $K, K' \in \preclustering_C$, we have
    \[
        W(K,K',N_1(K)\cap N_1(K'))  > \varepsilon \cdot (d(K)+d(K')).
    \]
\end{lemma}
\begin{proof}
    By Lemma~\ref{lem:degree-similar} and the definition of $N_1(\cdot)$, we can derive that $\preclustering_C \subseteq N_1(K) \cap N_1(K')$.
    By the definition of $W(\cdot, \cdot, \cdot)$, we have $W(K, K', N_1(K) \cap N_1(K') ) \geq W(K, K', \preclustering_C)$.
    Hence, it suffices to show that $W(K, K', \preclustering_C) > \varepsilon \cdot (d(K) + d(K'))$.
    We need the following simple fact.
    \begin{fact}\label{fact:mult-bound}
        $xy \ge x+y-1$ for $x,y\in [0,1]$.
    \end{fact}
    \begin{proof}
        Since $1-x \ge 0$ and $1-y \ge 0$, $(1-x)(1-y)=1-(x+y)+xy \ge 0$.
    \end{proof}
    Using this fact, we can obtain
    \begin{align*}
        W(K,K',\preclustering_C) 
        & = \sum_{K'' \in \preclustering_C \setminus \{K,K'\}} \left[ |K''|\cdot\frac{w^+(K,K'')}{|K|\cdot |K''|}\cdot \frac{w^+(K',K'')}{|K'|\cdot |K''|} \right]
        \\&
        \qquad +\frac{w^+(K,K')}{|K|}+\frac{w^+(K',K)}{|K'|}
        \\&
        \ge \sum_{K'' \in \preclustering_C \setminus \{K,K'\}}\left[\frac{w^+(K,K'')}{|K|}+ \frac{w^+(K',K'')}{|K'|}-|K''| \right] 
        \\&
        \qquad + \frac{w^+(K,K')}{|K|} +\frac{w^+(K',K)}{|K'|}
        \\&
        = \frac{w^+(K,C\setminus K)}{|K|} + \frac{w^+(K',C\setminus K')}{|K'|} - |C| + |K| + |K'|
        \\&
        =\frac{w^+(K,C\setminus K)}{|K|}-\frac{|C\setminus K|}{2}+\frac{|K|}{2}
        \\&
        \qquad +\frac{w^+(K',C\setminus K')}{|K'|}-\frac{|C\setminus K'|}{2}+\frac{|K'|}{2}
        \\&
        > \varepsilon \cdot (d(K)+d(K')),
    \end{align*}
    where the first inequality is due to Fact~\ref{fact:mult-bound}, and the last inequality is due to Inequality~\eqref{eq:atom-degree-ub-2}.
\end{proof}

For a precluster $K \in \preclustering$, let $N_2(K):=\{K' \in \preclustering \setminus \{K\} \mid K,K' \in N_1(K)\cap N_1(K')\}$.
Moreover, for two distinct preclusters $K, K' \in \preclustering$, we denote by $K \sim K'$ if
\begin{itemize}
    \item $K'\in N_2(K)$ and
    \item $W(K,K',N_1(K)\cap N_1(K'))>\varepsilon(d(K)+d(K'))$.
\end{itemize}
We are now ready to define the admissible edges; let 
\[
    \Eadm:=\{ uv\in E \mid  \exists K,K'\in\preclustering \text{ such that } K\sim K', u\in K, v\in K' \}.
\]
Observe that the definition of $\Eadm$ is independent from the nearly optimal solution $(\scriptCstarnear, \chistarnear)$.
However, we can derive from Lemma~\ref{lem:preclu:W} that $(\scriptCstarnear, \chistarnear)$ respects the constructed preclustered instance $(\preclustering, \colorpreclustering, \Eadm)$.
It remains to show that the size of $\Eadm$ is small enough.
\begin{lemma}\label{lem:adm-edges-bound}
    $|\Eadm| \le O(\varepsilon^{-2}) \, \opt$.
\end{lemma}
\begin{proof}
    Due to Lemma~\ref{lem:precluster-constant-approx} (or Lemma~\ref{lem:precluster-constant-approx_weighted} for \WCCC), it suffices to prove 
    \[
    |\Eadm| \le O(\varepsilon^{-2})\, \obj(\preclustering,\colorpreclustering), 
    \]
    where $\obj(\preclustering, \colorpreclustering)$ denotes the cost of the preclustering $(\preclustering, \colorpreclustering)$.
    To this end, we show that, for any $K\in \preclustering$,
    \begin{equation} \label{ineq:preclu:Eadm01}
        |K| \cdot \left( \sum_{K'\in \preclustering:K\sim K'} |K'| \right)
        \le O(\varepsilon^{-2})\, w^+(K,V \setminus K).
    \end{equation}
    If the above inequality is true, we immediately prove the lemma since
    \begin{align*}
        |\Eadm|
        & 
        \leq \sum_{K \in \preclustering} |K| \cdot \left( \sum_{K'\in \preclustering:K\sim K'} |K'| \right)
        \\&
        \leq O(\varepsilon^{-2}) \, \sum_{K \in \preclustering} w^+(K, V \setminus K)
        \\&
        \leq O(\varepsilon^{-2}) \, \obj(\preclustering, \colorpreclustering).
    \end{align*}

    Observe first that the left-hand side of Inequality~\eqref{ineq:preclu:Eadm01} is bounded by
    \begin{align}
        &
        |K| \left( \sum_{K'\in \preclustering:K\sim K'}|K'| \right)
        \nonumber \\&
        = |K| \left( \sum_{K'\in N_2(K)}|K'| \cdot \I \left[ W(K,K',N_1(K)\cap N_1(K'))>\varepsilon(d(K)+d(K')) \right] \right)
        \nonumber\\&
        \le |K| \left( \sum_{K'\in N_2(K)}|K'|\cdot\I \left[ W(K,K',N_1(K)\cap N_1(K'))>\varepsilon\cdot d(K) \right] \right)
        \nonumber\\&
        \le |K| \left( \sum_{K'\in N_2(K)}|K'|\cdot \frac{W(K,K',N_1(K)\cap N_1(K'))}{\varepsilon\cdot d(K)} \right)
        \nonumber\\&
        = \frac{|K|}{\varepsilon\cdot d(K)} \left( \sum_{K'\in N_2(K)}|K'|\cdot W(K,K',N_1(K)\cap N_1(K'))\label{eq:adm-bound-1} \right).
    \end{align}
    We further expand the summation of the right-hand side of Inequality~\eqref{eq:adm-bound-1} as follows:
    \begin{align*}
        &
        \sum_{K'\in N_2(K)}|K'|\cdot W(K,K',N_1(K)\cap N_1(K')) 
        \nonumber \\&
        = \sum_{K'\in N_2(K)}|K'|\cdot \Bigg(\sum_{\substack{K'' \in N_1(K)\cap N_1(K') : \\ K''\notin \{K,K'\}}}|K''|\cdot\left(\frac{w^+(K,K'')}{|K|\cdot |K''|}\cdot \frac{w^+(K',K'')}{|K'|\cdot |K''|}\right)\Bigg)
        \nonumber\\& \qquad
        + \sum_{K'\in N_2(K)}|K'|\cdot \left(\frac{w^+(K,K')}{|K|}+\frac{w^+(K',K)}{|K'|}\right) 
        \nonumber \\&
        = \underbrace{\sum_{K'\in N_2(K)} \Bigg(\sum_{\substack{K'' \in N_1(K)\cap N_1(K') : \\ K''\notin \{K,K'\}}}\frac{w^+(K,K'')}{|K|\cdot |K''|}\cdot w^+(K',K'') \Bigg)}_{\text{(I)}} 
        \\& \qquad 
        + \underbrace{\sum_{K'\in N_2(K)}\frac{w^+(K,K')}{|K|\cdot |K'|}\cdot|K'|^2}_{\text{(II)}}
        \; + \; \underbrace{\sum_{K'\in N_2(K)}w^+(K',K)}_{\text{(III)}}. 
    \end{align*}
    
    Let us first bound (I) and (II) from above.
    We can rearrange the order of summation in (I) as follows:
    \begin{align}
        \text{(I)} 
        & = \sum_{K'\in N_2(K)} \Bigg(\sum_{\substack{K'' \in N_1(K)\cap N_1(K') : \\ K''\notin \{K,K'\}}}\frac{w^+(K,K'')}{|K|\cdot |K''|}\cdot w^+(K',K'') \Bigg) 
        \nonumber \\ &
        =\sum_{\substack{K''\in N_1(K) :\\ K''\neq K}} \Bigg( \sum_{\substack{K' \in N_2(K) : \\ K'' \in N_1(K)\cap N_1(K') \\ K'\neq K''}}\frac{w^+(K,K'')}{|K|\cdot |K''|}\cdot w^+(K',K'') \Bigg)
        \nonumber\\&
        \le \sum_{\substack{K''\in N_1(K) :\\ K''\neq K}} \frac{w^+(K,K'')}{|K|\cdot |K''|} \cdot w^+(K'',V\setminus K'').\label{eq:adm-bound-3-1-1}
    \end{align}
    Hence, the summation of (I) and (II) is further bounded by
    \begin{align}
        &
        \text{(I)} + \text{(II)}
        \nonumber \\&
        \leq \sum_{\substack{K''\in N_1(K) :\\ K''\neq K}} \frac{w^+(K,K'')}{|K|\cdot |K''|} \cdot w^+(K'',V\setminus K'')+\sum_{K'\in N_2(K)}\frac{w^+(K,K')}{|K|\cdot |K'|}\cdot|K'|^2
        \nonumber \\&
        \le \sum_{\substack{K''\in N_1(K) :\\ K''\neq K}} \frac{w^+(K,K'')}{|K|\cdot |K''|}\cdot\left( w^+(K'',V\setminus K'') +|K''|^2\right)
        \nonumber\\&
        \le \sum_{\substack{K''\in N_1(K) :\\ K''\neq K}} \frac{w^+(K,K'')}{|K|}\cdot 2d(K'') 
        < \sum_{\substack{K''\in N_1(K) :\\ K''\neq K}} \frac{w^+(K,K'')}{|K|}\cdot \frac{2}{\varepsilon} d(K) 
        \nonumber\\&
        \le \frac{2}{\varepsilon} \cdot \frac{w^+(K,V\setminus K)}{|K|}\cdot d(K),\label{eq:adm-bound-3-1}
    \end{align}
    where the first inequality comes from Inequality~\eqref{eq:adm-bound-3-1-1}, the second inequality comes from the fact that $N_2(K) \subseteq N_1(K)\setminus \{K\}$, the third from the definition of $d(\cdot)$, and the fourth from the fact that $K'' \in N_1(K)$. 
    Lastly, we upper-bound (III) as follows:
    \begin{align}
        \text{(III)} 
        = \sum_{K'\in N_2(K)}w^+(K',K) 
        \le w^+(K,V\setminus K). \label{eq:adm-bound-3-2}
    \end{align}
    
    Plugging Inequalities~\eqref{eq:adm-bound-3-1} and~\eqref{eq:adm-bound-3-2} to Inequality~\eqref{eq:adm-bound-1} gives
    \begin{align*}
        &
        |K| \left( \sum_{K' \in \preclustering : K \sim K'} |K'| \right)
        \\&
        \leq \frac{|K|}{\varepsilon\cdot d(K)} \cdot \left(\frac{2}{\varepsilon} \cdot \frac{w^+(K,V\setminus K)}{|K|} \cdot d(K) + w^+(K,V\setminus K)\right) 
        \\& 
        =\left(\frac{2}{\varepsilon^2} + \frac{|K|}{\varepsilon\cdot d(K)} \right) w^+(K,V\setminus K) 
        \\&
        \le \left(\frac{2}{\varepsilon^2} + \frac{2}{\varepsilon} \right) w^+(K,V\setminus K),
    \end{align*}
    where the inequality is derived from the fact that $d(K) \ge \frac{|K|}{2}$, completing the proof of Inequality~\eqref{ineq:preclu:Eadm01}.
\end{proof}

\section{Deferred analysis from Section~\ref{sec_sampling}}\label{app:sampling}
\subsection{Bounded sub-cluster LP}

We first restrict the search space as follows:
\begin{lemma} \label{lem:searchspace}
    There exists a solution respecting $(\preclustering, \colorpreclustering, \Eadm)$ whose cost is still within $(1 + O(\varepsilon)) \, \opt$ while satisfying that, for every precluster $K \in \preclustering$, $K$ is either a cluster itself or a subset of a cluster of size greater than $|K| + \varepsilon_1 |\Nadm(K)|$ in the solution.
\end{lemma}
\begin{proof}
    Starting from the solution guaranteed by Theorem~\ref{thm:preclu:main}, consider the process that, while there exists a precluster $K$ contained in a cluster $C$ of size $|K| < |C| \leq |K| + \varepsilon_1 |\Nadm(K)|$, splits $C$ into $K$ and $C \setminus K$ with the same color.
    Observe that, whenever $C$ is split into $K$ and $C \setminus K$, the solution cost increases by at most
    \[
        |K| (|C| - |K|)
        \leq |K| \varepsilon_1 |\Nadm(K)|
        = \varepsilon_1 \sum_{u \in K} |\Nadm(u)|.
    \]
    Once a precluster is split, it is never processed anymore, so the total increase is bounded by 
    \[
        \varepsilon_1 \sum_{u \in V} |\Nadm(u)| = 2\varepsilon_1 |\Eadm| \leq O(\varepsilon) \opt,
    \]
    where the inequality is due to the definition of $\varepsilon_1 = \varepsilon^3$ and the upper bound of $|\Eadm|$ in Theorem~\ref{thm:preclu:main}.
\end{proof}

For color $\ell \in L$ and size $s \in [n]$, let $y^{\ell, s}_S$ indicate whether $S$ is a subset of a cluster of size $s$ colored $\ell$.
For a large enough integer $r = \Theta(\varepsilon^{-12})$, we define the \emph{bounded sub-cluster LP} as follows:
\begin{align}
    \text{min } & \obj(x) \nonumber 
    \\
    \text{s.t. }
    & \textstyle y^{\ell}_S = \sum_{s=1}^n y^{\ell, s}_S,
    && \forall \ell \in L \; \forall S \subseteq V : |S| \leq r, \label{const:bsclp:01} 
    \\
    & t^\ell_v = 1 - y^\ell_v,
    && \forall \ell \in L \; \forall v \in V, \label{const:bsclp:02}
    \\
    & x^\ell_{uv} = 1 - y^\ell_{uv}, 
    && \forall \ell \in L \; \forall uv \in \textstyle\binom{V}{2}, \label{const:bsclp:03} 
    \\
    & \textstyle \sum_{\ell \in L} y^\ell_v = 1,
    && \forall v \in V, \label{const:bsclp:04} 
    \\
    & \textstyle \sum_{v \in V} y^{\ell, s}_{Sv} = s \; y^{\ell, s}_S, 
    && \forall \ell \in L \; \forall s \in [n] \; \forall S \subseteq V : |S| \leq r-1, \label{const:bsclp:05}
    \\
    & x^{\colorpreclustering(K)}_{uv} = 0, 
    && \forall uv \subseteq K \in \preclustering_{>1}, \label{const:bsclp:06} 
    \\
    & x^{\ell}_{uv} = 1, 
    && \forall uv \subseteq K \in \preclustering_{>1} \; \forall \ell \neq \colorpreclustering(K), \label{const:bsclp:07} 
    \\
    & y^{\ell, s}_{uv} = y^{\ell, s}_u,
    && \forall \ell \in L \; \forall s \in [n] \; \forall u \in V \; \forall v \in K_u \label{const:bsclp:0705} 
    \\
    & x^{\ell}_{uv} = 1,
    && \forall \text{non-admissible $uv$} \; \forall \ell \in L, \label{const:bsclp:08} 
    \\
    & y^{\ell, s}_v = 0, 
    \nonumber \\
    & \mathrlap{\hspace{5em} \forall \ell \in L \; \forall v \in V \; \forall s \in [|K_v| - 1] \cup [|K_v| + 1, |K_v| + \varepsilon_1 |\Nadm(v)|], }
    \label{const:bsclp:09} 
    \\
    & y^{\colorpreclustering(K), |K|}_S = y^{\colorpreclustering(K), |K|}_v, 
    && \forall v \in K \in \preclustering_{>1} \; \forall S \subseteq K : |S| \leq r, \label{const:bsclp:10} 
    \\
    & y^{\colorpreclustering(K), |K|}_S = 0,
    && \forall K \in \preclustering_{>1} \; \forall S \not\subseteq K : |S| \leq r, S\cap K \neq \emptyset, \label{const:bsclp:11} 
    \\
    & \textstyle\sum_{T' \subseteq T} (-1)^{|T'|} y^{\ell, s}_{S \cup T'} \in [0, y^{\ell, s}_S],
    \nonumber \\
    & \mathrlap{\hspace{5em} \forall \ell \in L \; \forall s \in [n] \; \forall S, T \subseteq V : S \cap T = \emptyset, |S \cup T| \leq r,} \label{const:bsclp:12}
    \\
    & y^{\ell, s}_S \geq 0,
    && \forall \ell \in L \; \forall s \in [n] \; \forall S \subseteq V : |S| \leq r, \label{const:bsclp:13}
\end{align}
We remark that the difference between \CCC and \WCCC is only in the objective function, $\obj(x)$.
We now describe the constraints:
\eqref{const:bsclp:01} defines $y^\ell_S$ as the variable indicating if $S$ is a subset of a cluster colored $\ell$.
\eqref{const:bsclp:02} and \eqref{const:bsclp:03} correspond to the definitions of $t^\ell_v$ and $x^\ell_{uv}$ in the chromatic cluster LPs, respectively.
\eqref{const:bsclp:04} requires that each vertex $v \in V$ must belong to exactly one cluster.
\eqref{const:bsclp:05} indicates that, if $S \subseteq V$ is a subset of a cluster of size $s$, there exist exactly $s$ vertices $v$ (possibly in $S$) such that $S \cup \{v\}$ is a subset of a same cluster.
\eqref{const:bsclp:06} and \eqref{const:bsclp:07} together indicate that every vertex in a non-singleton precluster $K \in \preclustering_{>1}$ should be in a same cluster of color $\colorpreclustering(K)$ while \eqref{const:bsclp:0705} describes that, for any $u \in V$, every $v \in K_u$ must be in a same cluster.
\eqref{const:bsclp:08} restricts a solution to respect the non-admissible edges.
\eqref{const:bsclp:09}, \eqref{const:bsclp:10}, and \eqref{const:bsclp:11} are due to Lemma~\ref{lem:searchspace}.
\eqref{const:bsclp:12} is for the Sherali-Adams hierarchy, and \eqref{const:bsclp:13} is the nonnegativity constraint.

Since the size of the bounded sub-cluster LP is indeed bounded by $n^{\mathsf{poly}(\varepsilon^{-1})}$, we can solve this LP in time $n^{\mathsf{poly}(\varepsilon^{-1})}$.
\subsection{Sampling one cluster}
Let $y$ be an optimal solution to the bounded sub-cluster LP.
We sample one cluster from $y$ in the following manner.
We first randomly sample a color $\ell \in L$ with probability $\frac{y^\ell_\emptyset}{y_\emptyset}$, where $y_\emptyset := \sum_{\ell' \in L} y^{\ell'}_\emptyset$.
We then randomly sample a cardinality $s \in [n]$ with probability $\frac{y^{\ell, s}_\emptyset}{y^\ell_\emptyset}$.
Lastly, we sample a pivot $u \in V$ with probability $\frac{y^{\ell, s}_u}{s y^{\ell, s}_\emptyset}$.
With these sampled $\ell$, $s$, and $u$, we define $y'$ as $y'_S := \frac{y^{\ell, s}_{Su}}{y^{\ell, s}_{u}}$ for every $S \subseteq V$ of size $|S| \leq r-1$.
We now run Raghavendra and Tan's rounding algorithm~\cite{raghavendra2012approximating} upon $y'$ to sample a cluster $C \subseteq V$, and return $C$ colored $
\chi(C) := \ell$.

We now analyze this sampling algorithm.
Due to Raghavendra and Tan~\cite{raghavendra2012approximating}, we have:
\begin{lemma}[\cite{raghavendra2012approximating}] \label{lem:raghatan}
    One can sample $C \subseteq V$ in time $n^{O(r)}$ such that:
    \begin{itemize}
        \item for each $v \in V$, $\Pr[v \in C] = y'_v$;
        \item $\displaystyle \sum_{v, w \in \Nadm(u)} |\Pr[vw \subseteq C] - y'_{vw}| \leq \epsrt\cdot |\Nadm(u)|^2$.
    \end{itemize}
\end{lemma}

The probability of a vertex being in a cluster is as follows.
\begin{lemma} \label{lem:sampleone:color}
    For any $v \in V$ and color $\ell \in L$, $\Pr[\chi(C) = \ell \wedge v \in C] = \frac{y^\ell_v}{y_\emptyset}$.
\end{lemma}
\begin{proof}
    Observe that the probability is
    \begin{align*}
        \Pr[\chi(C) = \ell \wedge v \in C]
        &
        = \frac{y^\ell_\emptyset}{y_\emptyset} \sum_{s \in [n]} \frac{y^{\ell, s}_\emptyset}{y^\ell_\emptyset} \sum_{u \in V} \frac{y^{\ell, s}_{u}}{s y^{\ell, s}_\emptyset} \cdot \frac{y^{\ell, s}_{uv}}{y^{\ell, s}_u}
        \\ &
        = \frac{1}{y_\emptyset} \sum_{s \in [n]} \frac{1}{s} \sum_{u \in V} y^{\ell, s}_{uv} 
        = \frac{1}{y_\emptyset} \sum_{s \in [n]} y^{\ell, s}_v
        = \frac{y^{\ell}_v}{y_\emptyset},
    \end{align*}
    where the last two equalities are due to \eqref{const:bsclp:05} and \eqref{const:bsclp:01}, respectively.
\end{proof}

\begin{corollary} \label{cor:sampleone}
We have the following:
    \begin{enumerate}
        \item For any $v \in V$, $\Pr[v \in C] = \frac{1}{y_\emptyset}$.
        \item For any $v \in V$ and color $\ell \in L$, $\Pr[\chi(C) \neq \ell \wedge v \in C] = \frac{t^\ell_v}{y_\emptyset}$.
    \end{enumerate}    
\end{corollary}
\begin{proof}
    The first statement holds because
    \[
        \Pr[v \in C] = \sum_{\ell \in L} \Pr[\chi(C) = \ell \wedge v \in C] = \sum_{\ell \in L} \frac{y^\ell_v}{y_\emptyset} = \frac{1}{y_\emptyset},
    \]
    where the last equality comes from \eqref{const:bsclp:04}.
    The second statement is also easy to derive from \eqref{const:bsclp:02}.
\end{proof}

To bound the probabilities that two vertices are \emph{contained in} a same cluster, we define $\err^{\ell, s}_{vw|u}$ as the error generated by the procedure conditioned on color $\ell$, cardinality $s$, and pivot $u$. That is,
\[
    \err^{\ell, s}_{vw|u} := \left| \Pr[vw \subseteq C \mid \ell, s, u] - \frac{y^{\ell, s}_{uvw}}{y^{\ell, s}_u} \right| \; \text{for all $vw \in \binom{V}{2}$.}
\]
We further define
\begin{equation*}
    \err^{\ell, s}_{vw} := \sum_{u \in V} \frac{y^{\ell, s}_u}{s y^{\ell, s}_\emptyset} \cdot \err^{\ell, s}_{vw|u}
    \quad\text{ and }\quad
    \err^{\ell}_{vw} := \sum_{s \in [n]} \frac{y^{\ell, s}_\emptyset}{y^\ell_\emptyset} \cdot \err^{\ell, s}_{vw}.
\end{equation*}

\begin{lemma} \label{lem:sampleone:edge}
    For any $vw \in \binom{V}{2}$ and color $\ell \in L$,
    \begin{enumerate}
        \item $\Pr[\chi(C) = \ell \wedge v \in C \wedge w \not\in C] \leq \frac{x^\ell_{vw} - t^\ell_v}{y_\emptyset} + \frac{y^\ell_\emptyset}{y_\emptyset} \, \err^\ell_{vw} = \frac{y^\ell_v - y^\ell_{vw}}{y_\emptyset} + \frac{y^\ell_\emptyset}{y_\emptyset} \, \err^\ell_{vw}  $ and \\ $\Pr[\chi(C) = \ell \wedge v \not\in C \wedge w \in C] \leq \frac{x^\ell_{vw} - t^\ell_w}{y_\emptyset} + \frac{y^\ell_\emptyset}{y_\emptyset} \, \err^\ell_{vw} = \frac{y^\ell_w - y^\ell_{vw}}{y_\emptyset} + \frac{y^\ell_\emptyset}{y_\emptyset} \, \err^\ell_{vw}$;
        \item $\Pr[\chi(C) = \ell \wedge vw \subseteq C] \leq \frac{y^\ell_{vw}}{y_\emptyset} + \frac{y^\ell_\emptyset}{y_\emptyset} \, \err^\ell_{vw}$.
    \end{enumerate}
\end{lemma}
\begin{proof}
    For the first statement, we show the first half; the second half can be shown symmetrically.
    Let us condition on color $\ell$ and cardinality $s \in [n]$.
    The event $\{v \in C \wedge w \not\in C\}$ happens if and only if $uv \subseteq C$ and $uvw \not\subseteq C$ for pivot $u$.
    We thus have
    \begin{align*}
        \Pr[v \in C \wedge w \not\in C \mid \ell, s]
        &
        \leq \sum_{u \in V} \frac{y^{\ell, s}_u}{s y^{\ell, s}_\emptyset} \, \left( \frac{y^{\ell, s}_{uv}}{y^{\ell, s}_u} - \frac{y^{\ell, s}_{uvw}}{y^{\ell, s}_u} + \err^{\ell, s}_{vw|u} \right)
        \\ &
        = \sum_{u \in V} \left( \frac{y^{\ell, s}_{uv}}{s y^{\ell, s}_\emptyset} - \frac{y^{\ell, s}_{uvw}}{sy^{\ell, s}_\emptyset} + \frac{y^{\ell, s}_u}{s y^{\ell, s}_\emptyset} \, \err^{\ell, s}_{vw|u} \right)
        \\ &
        = \frac{y^{\ell, s}_v}{y^{\ell, s}_\emptyset} - \frac{y^{\ell, s}_{vw}}{y^{\ell, s}_\emptyset} + \err^{\ell, s}_{vw},
    \end{align*}
    where the inequality follows from Lemma~\ref{lem:raghatan} and the definition of $\err^{\ell, s}_{vw|u}$.
    Unconditioning $s$ yields
    \begin{align*}
        \Pr[v \in C \wedge w \not\in C \mid \ell]
        &
        \leq \sum_{s \in [n]} \frac{y^{\ell, s}_\emptyset}{y^\ell_\emptyset} \left( \frac{y^{\ell, s}_v}{y^{\ell, s}_\emptyset} - \frac{y^{\ell, s}_{vw}}{y^{\ell, s}_\emptyset} + \err^{\ell, s}_{vw} \right)
        \\ & 
        = \frac{y^\ell_v}{y^\ell_\emptyset} - \frac{y^\ell_{vw}}{y^\ell_\emptyset} + \err^\ell_{vw}
        = \frac{x^\ell_{vw} - t^\ell_v}{y^\ell_\emptyset} + \err^\ell_{vw},
    \end{align*}
    where the last equality is derived from \eqref{const:bsclp:02} and \eqref{const:bsclp:03}.
    Multiplying both sides by $\Pr[\chi(C) = \ell] = \frac{y^\ell_\emptyset}{y_\emptyset}$ completes the proof of the first statement.

    The second statement can be shown in a similar way.
    Conditioned on color $\ell$, cardinality $s \in [n]$, and pivot $u \in V$, we have
    \[
        \Pr[vw \subseteq C \mid \ell, s, u] \leq \frac{y^{\ell, s}_{uvw}}{y^{\ell, s}_{u}} + \err^{\ell, s}_{vw|u}
    \]
    by the definition of $\err^{\ell, s}_{vw|u}$.
    The rest of the argument is analogous to the first statement.
\end{proof}

\begin{lemma} \label{lem:sumerror}
    \(
    \displaystyle \sum_{\ell \in L} \sum_{vw \in \binom{V}{2}}  y^\ell_\emptyset \err^\ell_{vw} \leq O(\varepsilon_1) |\Eadm|.
    \)
\end{lemma}
\begin{proof}
    Fix a color $\ell \in L$, a cardinality $s \in [n]$, and a pivot $u \in V$ with $y^{\ell, s}_u > 0$.
    If $s = 1$, the procedure samples $C = \{u\}$ where the errors are trivially 0.
    For $s \geq 2$, let $k_u := |K_u|$ be the size of the precluster containing $u$.
    Observe that $s$ is either $k_u$ or greater than $k_u + \varepsilon_1 |\Nadm(u)|$ due to \eqref{const:bsclp:09}.
    If $s = k_u$, the procedure samples $C = K_u$ due to \eqref{const:bsclp:10} and \eqref{const:bsclp:11}, so it is again easy to see that the errors are 0.
    
    We now consider the case where $s > k_u + \varepsilon_1 |\Nadm(u)|$.
    We show that, for any $vw \in \binom{V}{2}$, $\err^{\ell,s}_{vw|u}=0$ unless both $v$ and $w$ belong to $\Nadm(u)$.
    First, if one endpoint, say $v$, is $u$ itself, then
    \[
    \Pr[vw\subseteq C\mid \ell,s,u] = \Pr[w\in C\mid \ell,s,u] = y'_w = \frac{y^{\ell,s}_{uw}}{y^{\ell,s}_u} = \frac{y^{\ell,s}_{uvw}}{y^{\ell,s}_u}.
    \]
    Thus $\err^{\ell,s}_{vw|u}=0$.
    If an endpoint, say $v$, does not belong to $K_u \cup \Nadm(u)$, then $uv$ is neither atomic nor admissible, and hence is non-admissible by definition.
    This implies $y^{\ell,s}_{uv}=0$ by \eqref{const:bsclp:08}, \eqref{const:bsclp:03}, \eqref{const:bsclp:01}, and \eqref{const:bsclp:13}.
    By Lemma~\ref{lem:raghatan}, $\Pr[v\in C\mid \ell,s,u]=y'_v=0$.
    Moreover, $y^{\ell,s}_{uvw}=0$ by \eqref{const:bsclp:05} and \eqref{const:bsclp:13}.
    Therefore, $\err^{\ell,s}_{vw|u}=0$ whenever an endpoint is not in $K_u \cup \Nadm(u)$.
    It remains to consider the case where an endpoint, say $v$, belongs to $K_u\setminus\{u\}$.
    Note that $y^{\ell,s}_{uv}=y^{\ell,s}_u$ by \eqref{const:bsclp:0705}, and hence $\Pr[v\in C\mid \ell,s,u]=y'_v=1$ by Lemma~\ref{lem:raghatan}.
    Moreover, applying \eqref{const:bsclp:12} with $S={u,w}$ and $T={v}$ gives
    \[
    y^{\ell,s}_{uw}-y^{\ell,s}_{uvw}\ge 0.
    \]
    Applying the same constraint with $S={u}$ and $T={v,w}$ gives
    \[
    y^{\ell,s}_u-y^{\ell,s}_{uv}-y^{\ell,s}_{uw}+y^{\ell,s}_{uvw}\ge 0 \quad \Longleftrightarrow \quad y^{\ell,s}_{uw}-y^{\ell,s}_{uvw}\le y^{\ell,s}_u-y^{\ell,s}_{uv}.
    \]
    Since $y^{\ell,s}_{uv}=y^{\ell,s}_u$, the two inequalities above imply $y^{\ell,s}_{uvw}=y^{\ell,s}_{uw}$.
    Thus, pairs involving vertices of $K_u$ also incur no error.
    Therefore, only pairs within $\Nadm(u)$ can have nonzero error.

    We thus have
    \begin{align}
        \sum_{vw \in \binom{V}{2}} \err^{\ell, s}_{vw|u}
        &
        = \sum_{vw \subseteq \Nadm(u)} \err^{\ell, s}_{vw|u} 
        \leq \frac{1}{2} \, \sum_{v,w \in \Nadm(u)} \err^{\ell, s}_{vw|u} 
        \nonumber \\&
        \leq \frac{\epsrt}{2} \, |\Nadm(u)|^2
        \leq \frac{\varepsilon_1}{2} \, |\Nadm(u)| \, (s - k_u),
        \label{eq:sumerrlsvwu}
    \end{align}
    where the second inequality is due to Lemma~\ref{lem:raghatan} 
    and the last is due to the condition of $s$ and the definition of $\epsrt := \varepsilon_1^2$.
    We claim that 
    \(
        s - k_u = \sum_{v \in \Nadm(u)} \frac{y^{\ell, s}_{uv}}{y^{\ell, s}_u}.
    \)
    To see this, observe first that $y^{\ell, s}_{uv} = y^{\ell, s}_u$ for every $v \in K_u$ due to \eqref{const:bsclp:0705}.
    Therefore, from \eqref{const:bsclp:05}, we can derive
    \[
        s y^{\ell, s}_u 
        = \sum_{v \in V} y^{\ell, s}_{uv} 
        = \sum_{v \in K_u} y^{\ell, s}_{uv} + \sum_{v \in \Nadm(u)} y^{\ell, s}_{uv}
        = \sum_{v \in K_u} y^{\ell, s}_{u} + \sum_{v \in \Nadm(u)} y^{\ell, s}_{uv}.
    \]
    Dividing by $y^{\ell, s}_u$ and rearranging the terms completes the proof of the claim.
    We can thus further bound \eqref{eq:sumerrlsvwu} as follows:
    \[
        \sum_{vw \in \binom{V}{2}} \err^{\ell, s}_{vw|u}
        \leq \frac{\varepsilon_1}{2} |\Nadm(u)| \, \sum_{v \in \Nadm(u)} \frac{y^{\ell, s}_{uv}}{y^{\ell, s}_{u}}
        = \frac{\varepsilon_1}{2} \sum_{v \in \Nadm(u)} \sum_{w \in \Nadm(u)} \frac{y^{\ell, s}_{uv}}{y^{\ell, s}_{u}}.
    \]

    Unconditioning the pivot $u \in V$ yields
    \begin{align*}
        \sum_{vw \in \binom{V}{2}} \err^{\ell, s}_{vw}
        &
        \leq \sum_{u \in V} \frac{y^{\ell, s}_u}{s y^{\ell, s}_\emptyset} \, \frac{\varepsilon_1}{2} \sum_{v \in \Nadm(u)} \sum_{w \in \Nadm(u)} \frac{y^{\ell, s}_{uv}}{y^{\ell, s}_{u}}
        \\ &
        = \frac{\varepsilon_1}{2} \, \frac{1}{s y^{\ell, s}_\emptyset} \sum_{u \in V} \sum_{v \in \Nadm(u)} \sum_{w \in \Nadm(u)} y^{\ell, s}_{uv}
        \\ &
        = \frac{\varepsilon_1}{2} \, \frac{1}{s y^{\ell, s}_\emptyset} \sum_{v \in V} \sum_{u \in \Nadm(v)} \sum_{w \in \Nadm(u)} y^{\ell, s}_{uv}
        \\ &
        = \frac{\varepsilon_1}{2} \sum_{v \in V} \frac{y^{\ell, s}_{v}}{s y^{\ell, s}_\emptyset}  \sum_{u \in \Nadm(v)} \sum_{w \in \Nadm(u)} \frac{y^{\ell, s}_{uv}}{y^{\ell, s}_v}
        \\ &
        \stackrel{(a)}{\leq} \frac{\varepsilon_1}{2} \sum_{v \in V} \frac{y^{\ell, s}_{v}}{s y^{\ell, s}_\emptyset}  \sum_{uw \in \Eadm} \frac{y^{\ell, s}_{uv} + y^{\ell, s}_{vw}}{y^{\ell, s}_v}
        \\ &
        \stackrel{(b)}{\leq} \frac{\varepsilon_1}{2} \sum_{v \in V} \frac{y^{\ell, s}_{v}}{s y^{\ell, s}_\emptyset}  \sum_{uw \in \Eadm} 2 \Pr[C \cap \{u, w\} \neq \emptyset \mid \ell, s, v] 
        \\ &
        = \varepsilon_1 \sum_{uw \in \Eadm} \Pr[C \cap \{u, w\} \neq \emptyset \mid \ell, s],
    \end{align*}
    where $(a)$ follows from the fact that each admissible edge $u'w' \in \Eadm$ can contribute to the double summations at most twice --- once from $u = u'$ and $w = w'$ and the other time from $u = w'$ and $w = u'$; $(b)$ from the fact that $\frac{y^{\ell, s}_{uv}}{y^{\ell, s}_v} = \Pr[u \in C \mid \ell, s, v] \leq \Pr[C \cap \{u, w\} \neq \emptyset \mid \ell, s, v]$.

    We further obtain by unconditioning $s$
    \begin{align*}
        \sum_{vw \in \binom{V}{2}} \err^{\ell}_{vw}
        &
        \leq \sum_{s \in [n]} \frac{y^{\ell, s}_\emptyset}{y^\ell_\emptyset} \, \varepsilon_1 \sum_{uw \in \Eadm} \Pr[C \cap \{u, w\} \neq \emptyset \mid \ell, s] 
        \\ &
        = \varepsilon_1 \sum_{uw \in \Eadm} \Pr[ C \cap \{u, w\} \neq \emptyset \mid \ell].
    \end{align*}
    Therefore, we have
    \begin{align*}
        \sum_{\ell \in L} \frac{y^\ell_\emptyset}{y_\emptyset} \sum_{vw \in \binom{V}{2}} \err^\ell_{vw}
        &
        \leq \sum_{\ell \in L} \frac{y^\ell_\emptyset}{y_\emptyset} \cdot \varepsilon_1 \sum_{uw \in \Eadm} \Pr[C \cap \{u, w\} \neq \emptyset \mid \ell]
        \\ &
        = \varepsilon_1 \sum_{uw \in \Eadm} \sum_{\ell \in L} \Pr[\chi(C) = \ell \wedge C \cap \{u, w\} \neq \emptyset]
        \\ &
        \leq \varepsilon_1 \sum_{uw \in \Eadm} \sum_{\ell \in L} \left( \frac{y^\ell_u + y^\ell_w - y^\ell_{uw}}{y_\emptyset} + 3 \, \frac{y^\ell_\emptyset}{y_\emptyset} \err^\ell_{uw} \right)
        \\ &
        \leq \frac{2 \varepsilon_1}{y_\emptyset} |\Eadm| + 3 \varepsilon_1 \sum_{\ell \in L} \sum_{vw \in \binom{V}{2}} \frac{y^\ell_\emptyset}{y_\emptyset} \err^\ell_{vw},
    \end{align*}
    where the second inequality is satisfied due to Lemma~\ref{lem:sampleone:edge} while the last inequality follows from \eqref{const:bsclp:04}.
    Multiplying both sides by $y_\emptyset$ and rearranging the terms, we can derive
    \[
        \sum_{\ell \in L} \sum_{vw \in \binom{V}{2}} y^\ell_\emptyset \err^\ell_{vw} \leq \frac{2 \varepsilon_1}{1 - 3 \varepsilon_1} |\Eadm|
        = O(\varepsilon_1) |\Eadm|.
    \]
\end{proof}
\subsection{Monte Carlo sampling}
Recall $\varepsilon_1 = \varepsilon^3$.
We use the following version of Chernoff bound:
\begin{theorem}[Chernoff bound]\label{thm:chernoff}
    For independent Bernoulli random variables $X_1, \ldots, X_k \in \{0, 1\}$, let $X := \sum_{i \in [k]} X_i$, $\mu := \E[X]$, and $\mu' \geq \mu$ be a real number.
    Then, for any $\delta \in (0, 1)$, we have
    \begin{enumerate}
        \item $\displaystyle \Pr[X < (1 - \delta) \mu] < \exp \left(-\frac{\delta^2 \mu}{2} \right)$ and
        \item $\displaystyle \Pr[X > \mu + \delta \mu'] < \exp \left( -\frac{\delta^2 \mu'}{3} \right)$.
    \end{enumerate}
\end{theorem}

Let $\Delta := \Theta \left( \frac{n^4 \log n}{\varepsilon_1^2 |\Eadm|} \right)$ with a sufficiently large hidden constant.
(Here, we assume $\Eadm\neq\emptyset$; otherwise, $(\preclustering,\colorpreclustering)$ given by Theorem~\ref{thm:preclu:main}, with arbitrary colors assigned to singleton preclusters, is already a $(1+\varepsilon)$-approximate integral solution.)
In particular, we assume that $\Delta y_\emptyset$ is an integer and $\Delta \geq \frac{1}{\varepsilon - \varepsilon_1} = \frac{1}{\varepsilon - \varepsilon^3}$.
We then independently sample colored clusters from the previous procedure for $\Delta y_\emptyset$ times.
Let $C_1, \ldots, C_{\Delta y_\emptyset}$ be the sampled colored clusters.
Let $\ell_1, \ldots, \ell_{\Delta y_\emptyset}$ denote their corresponding colors.
For each vertex $v \in V$ and color $\ell \in L$, let $R^\ell_v := \{t \in \{1, 2, \ldots, \Delta y_\emptyset \} \mid \ell_t = \ell \wedge v \in C_t \}$.
We also define $R^{-\ell}_v := \bigcup_{\ell' \neq \ell} R^{\ell'}_v$ and $R_v := \bigcup_{\ell \in L} R^\ell_v$;
note that $R^{-\ell}_v = R_v \setminus R^\ell_v$.

By choosing a large enough constant for $\Delta$, we can ensure that, with high probability, the following events are all satisfied by Chernoff bound and the union bound.
\begin{itemize}
    \item For every $v \in V$, we have 
    \begin{equation} \label{eq:Rv}
        |R_v| \geq (1 - \varepsilon_1) \Delta.
    \end{equation}
    To see this, define $X_t := \I[v \in C_t]$, then by Corollary~\ref{cor:sampleone}, $\mu = \Delta $.
    Set $\delta := \varepsilon_1$.
    We then have
    \[
        \Pr[|R_v| < (1 - \varepsilon_1) \Delta] < \exp \left( - \frac{\varepsilon_1^2 \Delta}{2} \right) \leq n^{-\Theta(1)}
    \]
    from the first bound in Theorem~\ref{thm:chernoff}.
    \item For every $v \in V$ and $\ell \in L$, we have
    \begin{equation} \label{eq:Rmlv}
        |R^{-\ell}_v|
        \leq \Delta t^\ell_v + \varepsilon_1 \max \left\{ \Delta t^\ell_v, \frac{\Delta |\Eadm|}{n^4} \right\}
        \leq (1 + \varepsilon_1) \Delta t^\ell_v + \frac{\varepsilon_1 \Delta |\Eadm|}{n^4}. 
    \end{equation}
    Define $X_t := \I[\ell_t \neq \ell \wedge v \in C_t]$, and then by Corollary~\ref{cor:sampleone}, we have $\mu = \Delta t^\ell_v$.
    Let $\mu' := \max\{ \Delta t^\ell_v, \frac{\Delta |\Eadm|}{ n^4} \}$ and $\delta := \varepsilon_1$.
    The second bound in Theorem~\ref{thm:chernoff} implies
    \begin{align*}
        &
        \Pr \left[ |R^{-\ell}_v| > \Delta t^\ell_v + \varepsilon_1 \max \left\{ \Delta t^\ell_v, \frac{\Delta |\Eadm|}{n^4} \right\} \right]
        \\&
        \qquad \leq \exp \left(-\frac{\varepsilon_1^2 \frac{\Delta |\Eadm|}{n^4}}{3} \right)
        \\&
        \qquad \leq n^{-\Theta(1)}.
    \end{align*}
    \item For every $uv \in \binom{V}{2}$ and $\ell \in L$, we have
    \begin{align}
        &
        |R^\ell_u \setminus R^\ell_v| 
        \nonumber \\& 
        \leq \Delta (x^\ell_{uv} - t^\ell_u + y^\ell_\emptyset  \, \err^\ell_{uv}) + \varepsilon_1 \max \left \{ \Delta (x^\ell_{uv} - t^\ell_u + y^\ell_\emptyset  \, \err^\ell_{uv}), \frac{\Delta |\Eadm|}{n^4} \right \}
        \nonumber \\ & 
        \leq (1 + \varepsilon_1) \Delta (x^\ell_{uv} - t^\ell_u + y^\ell_\emptyset  \, \err^\ell_{uv}) + \frac{\varepsilon_1 \Delta |\Eadm|}{n^4}. \label{eq:RlumRlv}
    \end{align}
    Define $X_t := \I[\ell_t = \ell \wedge u \in C_t \wedge v \not\in C_t]$.
    By Lemma~\ref{lem:sampleone:edge}, we have $\mu \leq \Delta(x^\ell_{uv} - t^\ell_u + y^\ell_\emptyset \, \err^\ell_{uv})$.
    Let $\mu' := \max \left \{ \Delta (x^\ell_{uv} - t^\ell_u + y^\ell_\emptyset  \, \err^\ell_{uv}), \frac{\Delta |\Eadm|}{n^4} \right \}$ and $\delta := \varepsilon_1$.
    By the second bound, we obtain
    \begin{align*}
        &
        \Pr[|R^\ell_u \setminus R^\ell_v| > \Delta (x^\ell_{uv} - t^\ell_u + y^\ell_\emptyset  \, \err^\ell_{uv}) + \delta \mu']
        \\ & 
        \qquad \leq \Pr[|R^\ell_u \setminus R^\ell_v| > \mu + \delta \mu'] 
        \\ &
        \qquad \leq \exp \left( -\frac{\varepsilon_1^2 \frac{\Delta |\Eadm|}{n^4}}{3} \right)
        \\&
        \qquad \leq n^{-\Theta(1)}.
    \end{align*}
    A symmetric argument gives
    \begin{equation} \label{eq:RlvmRlu}
        |R^\ell_v \setminus R^\ell_u|
        \leq (1 + \varepsilon_1) \Delta (x^\ell_{uv} - t^\ell_v + y^\ell_\emptyset  \, \err^\ell_{uv}) + \frac{\varepsilon_1 \Delta |\Eadm|}{n^4}.
    \end{equation}
    Due to \eqref{eq:Rmlv}, \eqref{eq:RlumRlv}, and \eqref{eq:RlvmRlu}, we can derive
    \begin{align}
        |R_u| - |R^\ell_u \cap R^\ell_v|
        &
        = |R^{-\ell}_u| + |R^\ell_u \setminus R^\ell_v|
        \nonumber \\&
        \leq (1 + \varepsilon_1) \Delta (x^\ell_{uv} + y^\ell_\emptyset \err^\ell_{uv}) + \frac{2 \varepsilon_1 \Delta |\Eadm|}{n^4}; \label{eq:RumRlucapRlv}
        \\
        |R_v| - |R^\ell_v \cap R^\ell_u|
        & 
        = |R^{-\ell}_v| + |R^\ell_v \setminus R^\ell_u| 
        \nonumber \\&
        \leq (1 + \varepsilon_1) \Delta (x^\ell_{uv} + y^\ell_\emptyset \err^\ell_{uv}) + \frac{2 \varepsilon_1 \Delta |\Eadm|}{n^4}. \label{eq:RvmRlvcapRlu}
    \end{align}
    \item For every $uv \in \binom{V}{2}$ and $\ell \in L$, we have
    \begin{align}
        &
        |R^\ell_u \cap R^\ell_v| 
        \nonumber \\ & 
        \leq \Delta (y^\ell_{uv} + y^\ell_\emptyset  \, \err^\ell_{uv}) + \varepsilon_1 \max \left \{ \Delta (y^\ell_{uv} + y^\ell_\emptyset  \, \err^\ell_{uv}), \frac{\Delta |\Eadm|}{n^4} \right \}
        \nonumber \\ & 
        \leq (1 + \varepsilon_1) \Delta (y^\ell_{uv} + y^\ell_\emptyset  \, \err^\ell_{uv}) + \frac{\varepsilon_1 \Delta |\Eadm|}{n^4}. \label{eq:RlucapRlv}
    \end{align}
    Define $X_t := \I[\ell_t = \ell \wedge u \in C_t \wedge v \in C_t]$.
    By Lemma~\ref{lem:sampleone:edge}, we have $\mu \leq \Delta(y^\ell_{uv} + y^\ell_\emptyset \, \err^\ell_{uv})$.
    Let $\mu' := \max \left \{ \Delta (y^\ell_{uv} + y^\ell_\emptyset  \, \err^\ell_{uv}), \frac{\Delta |\Eadm|}{n^4} \right \}$ and $\delta := \varepsilon_1$.
    By the second bound, we obtain
    \begin{align*}
        &
        \Pr[|R^\ell_u \cap R^\ell_v| > \Delta (y^\ell_{uv} + y^\ell_\emptyset  \, \err^\ell_{uv}) + \delta \mu']
        \\ & 
        \qquad \leq \Pr[|R^\ell_u \cap R^\ell_v| > \mu + \delta \mu'] 
        \\ &
        \qquad \leq \exp \left( -\frac{\varepsilon_1^2 \frac{\Delta |\Eadm|}{n^4}}{3} \right)
        \\ & 
        \qquad \leq n^{-\Theta(1)}.
    \end{align*}
\end{itemize}

For every $v \in V$, let $\Rtilde_v$ be the set of the $\ceil{(1 - \varepsilon) \Delta}$ smallest indices in $R_v$; let $\Rtilde^{\ell}_v := R^\ell_v \cap \Rtilde_v$ and $\Rtilde^{-\ell}_v := R^{-\ell}_v \cap \Rtilde_v$.
For every $uv \in \binom{V}{2}$ and $\ell \in L$, it is easy to see that \eqref{eq:RlucapRlv} remains a valid upper bound since
\begin{equation} \label{eqRtlucapRtlv}
    |\Rtilde^\ell_u \cap \Rtilde^\ell_v| 
    \leq |R^\ell_u \cap R^\ell_v|
    \leq (1 + \varepsilon_1) \Delta (y^\ell_{uv} + y^\ell_\emptyset  \, \err^\ell_{uv}) + \frac{\varepsilon_1 \Delta |\Eadm|}{n^4}.
\end{equation}
We now argue that \eqref{eq:RumRlucapRlv} and \eqref{eq:RvmRlvcapRlu} are still satisfied.
\begin{lemma} \label{lem:RtumRtlucapRtlv}
    For every $uv \in \binom{V}{2}$ and $\ell \in L$, we have
    \begin{align*}
        |\Rtilde_u| - |\Rtilde^\ell_u \cap \Rtilde^\ell_v|
        &
        = |\Rtilde_v| - |\Rtilde^\ell_u \cap \Rtilde^\ell_v|
        \\&
        \leq (1 + \varepsilon_1) \Delta (x^\ell_{uv} + y^\ell_\emptyset \err^\ell_{uv}) + \frac{2 \varepsilon_1 \Delta |\Eadm|}{n^4}.
    \end{align*}
\end{lemma}
\begin{proof}
    The equality is trivial since $|\Rtilde_u| = |\Rtilde_v| = \ceil{(1 - \varepsilon) \Delta}$.
    To see the inequality, imagine the following process of removing indices from $R_u$ and $R_v$ that eventually outputs $\Rtilde_u$ and $\Rtilde_v$, respectively.
    Assume without loss of generality that $|R_u| \geq |R_v|$ at the very beginning; we thus have
    \begin{equation} \label{eq:removeinv}
        |R_v| - |R^\ell_v \cap R^\ell_u| \leq |R_u| - |R^\ell_u \cap R^\ell_v| \leq \tau, 
    \end{equation}
    where $\tau := (1 + \varepsilon_1) \Delta (x^\ell_{uv} + y^\ell_\emptyset \err^\ell_{uv}) + \frac{2 \varepsilon_1 \Delta |\Eadm|}{n^4}$.
    Until $|R_u| = |R_v|$, we remove the largest index from $R_u$.
    Note that $|R_u| - |R^\ell_u \cap R^\ell_v|$ either stays the same if the removed index is in $R^\ell_u \cap R^\ell_v$ or decreases by one otherwise.
    As we still have $|R_u| \geq |R_v|$ after the removal, \eqref{eq:removeinv} is still satisfied.
    
    Once we reach the point of $|R_u| = |R_v| > \ceil{(1-\varepsilon) \Delta}$, we have
    \(
        |R_u| - |R^\ell_u \cap R^\ell_v| = |R_v| - |R^\ell_v \cap R^\ell_u| \leq \tau.
    \)
    From this point, we remove the largest index from each of $R_u$ and $R_v$, respectively, until $|R_u| = |R_v| = \ceil{(1-\varepsilon) \Delta}$.
    It is easy to see that, at each step, $|R_u|$ and $|R_v|$ decrease exactly by one, respectively.
    We claim that $|R^\ell_u \cap R^\ell_v|$ decreases by at most one.
    Indeed, to have $|R^\ell_u \cap R^\ell_v|$ decrease by two, we must remove two distinct indices $i, j \in R^\ell_u \cap R^\ell_v$ from $R_u$ and $R_v$, respectively; however, since both $i$ and $j$ belong to both $R_u$ and $R_v$, this contradicts the choice of index to be removed from either $R_u$ or $R_v$.
    Due to this claim, we can see that $|R_u| - |R^\ell_u \cap R^\ell_v|$ and $|R_v| - |R^\ell_v \cap R^\ell_u|$ never increase over each step of this process, implying the lemma.
\end{proof}

For every $t \in \{1, 2, \ldots, \Delta y_\emptyset \}$, we define $\Ctilde_t := \{ v \in V \mid t \in \Rtilde_v \} \subseteq C_t$.
Recall $\ell_t$ denotes the color of $C_t$.
Let $\ztilde^\ell_S := \frac{|\{ t \mid \Ctilde_t = S \wedge \ell_t = \ell \}|}{\ceil{(1 - \varepsilon) \Delta}}$ for every $\ell \in L$ and $S \subseteq V$.
Observe that, for every $v \in V$,
\[
    \sum_{S \ni v, \ell \in L} \ztilde^\ell_S
    = \frac{|\Rtilde_v|}{\ceil{(1 - \varepsilon) \Delta}} 
    = 1,
\]
showing that $\ztilde$ is feasible to the chromatic cluster LP.

To upper bound the objective value of this solution, for every $uv \in \binom{V}{2}$ and $\ell \in L$, let $\xtilde^\ell_{uv} := 1 - \sum_{S \supseteq uv} \ztilde^\ell_S$.
Note that we have
\begin{align*}
    \xtilde^\ell_{uv}
    & = \frac{|\Rtilde_u| - |\Rtilde^\ell_u \cap \Rtilde^\ell_v|}{\ceil{(1 - \varepsilon) \Delta}}
    \leq \frac{1 + \varepsilon_1}{1 - \varepsilon} (x^\ell_{uv} + y^\ell_\emptyset \err^\ell_{uv}) + \frac{2 \varepsilon_1 |\Eadm|}{(1 - \varepsilon) n^4}; \\
    1 - \xtilde^\ell_{uv}
    & = \frac{|\Rtilde^\ell_u \cap \Rtilde^\ell_v|}{\ceil{(1 - \varepsilon) \Delta}}
    \leq \frac{1 + \varepsilon_1}{1 - \varepsilon} (1 - x^\ell_{uv} + y^\ell_\emptyset \err^\ell_{uv}) + \frac{\varepsilon_1 |\Eadm|}{(1 - \varepsilon) n^4},
\end{align*}
where the first half is due to Lemma~\ref{lem:RtumRtlucapRtlv} and the second half is due to \eqref{eqRtlucapRtlv} and \eqref{const:bsclp:03}.
We thus have
\begin{align*}
    \obj(\xtilde)
    &
    \leq (1 + O(\varepsilon)) \obj(x) + \sum_{uv \in E} \sum_{\ell \in L} \left[ (1 + O(\varepsilon)) \, y^\ell_\emptyset \err^\ell_{uv} + O(\varepsilon_1) \, \frac{|\Eadm|}{n^4} \right]
    \\ &
    \leq (1 + O(\varepsilon)) \obj(x) + O(\varepsilon_1) |\Eadm|
    \\ &
    \leq (1 + O(\varepsilon)) \opt + O(\varepsilon^3) \, O(\varepsilon^{-2}) \, \opt
    \\ &
    \leq (1 + O(\varepsilon)) \opt,
\end{align*}
where the second inequality is due to Lemma~\ref{lem:sumerror} and the fact that $|L| \leq n^2$ and the third inequality is due to Lemma~\ref{lem:searchspace} and Theorem~\ref{thm:preclu:main}.

\end{document}